\documentclass[ aps,prb,amsmath,amssymb,superscriptaddress, showpacs, twoside]{revtex4}

\input {epsf.sty}
\usepackage{graphicx}
\usepackage{epsfig}
\newcommand{\be}[1]{ \begin{eqnarray} \mbox{$\label{#1}$} }

\newcommand{\ee}{\end{eqnarray}}
\newcommand{\pref}[1]{(\ref{#1})}
\newcommand{\oncite}[1]{Ref. \onlinecite{#1} }
\newcommand{\oncitep}[1]{Ref. \onlinecite{#1}. }
\newcommand{\oncitec}[1]{Ref. \onlinecite{#1}, }

\newcounter{mycount}

\newcommand\ie {{\it i.e. }}
\newcommand\eg {{\it e.g. }}

\newcommand\cf {{\it cf.  }}
\newcommand\etal{{\it et al.} }

\newcommand\half{\frac 1 2 }

\newcommand\Pf{\mbox{Pf}}

\newcommand\ket [1] {|#1 \rangle }
\newcommand\bra [1] {\langle #1 |} 
\newcommand{\bracket}[2]   {  \left<#1 |  #2\right>}
\newcommand{\av}[1]{\langle #1\rangle}

\newcommand\noi{\noindent}

\begin{document}

\title{ Quantum Hall quasielectron operators in conformal field theory}
\author{T.H. Hansson}
\author{M. Hermanns} 
\affiliation {Department of Physics, Stockholm University
AlbaNova University Center,
SE - 106 91 Stockholm, Sweden} 
\author{ S. Viefers}
\affiliation{ Department of Physics, University of Oslo, P.O. Box 1048 Blindern, 0316 Oslo, Norway}

\date{\today}

\begin{abstract} 

In the conformal field theory (CFT) approach to the quantum Hall effect, the multi-electron wave functions are expressed as correlation functions in certain rational CFTs. While this approach has led to a well-understood description of the  fractionally charged  quasihole excitations, the quasielectrons have turned out to be much harder to handle. In particular, forming quasielectron states requires  non-local operators, in sharp contrast to quasiholes that can be created by local chiral vertex operators. In both cases, the operators are strongly constrained by general requirements of symmetry, braiding and fusion. Here we construct a quasielectron operator satisfying these demands and show that it reproduces known good quasiparticle wave functions, as well as predicts new ones. In particular we propose explicit wave functions for quasielectron excitations of the Moore-Read Pfaffian state. 
Further, this operator allows us to explicitly express the composite fermion wave functions in the positive Jain series in hierarchical form, thus settling a longtime controversy. We also critically discuss the status of the fractional statistics of quasiparticles in the Abelian hierarchical quantum Hall states, and argue that our construction of  localized quasielectron states sheds new light on their statistics.  
At the technical level we introduce  a generalized normal ordering, that allows us to "fuse" an electron operator with the inverse of an hole operator, and also an alternative approach to the background charge needed to neutralize CFT correlators. As a result we get a fully holomorphic CFT representation of a large set of quantum Hall wave functions.

\end{abstract}
\pacs{73.43.Cd, 11.25.Hf, 71.10.Pm}

\maketitle

\newcommand\jas[3]{(z_{#1} - z_{#2})^{#3}}
\newcommand\prs[2] {\prod_{#1}\!^{(#2)}}
\newcommand\pr[3] {\prod_{#1<#2}\!^{(#3)}}
\newcommand\prjas[4] {\pr #1 #2 {#3}  \jas #1 #2 #4}
\newcommand\vmea[1] {e^{i\sqrt{m} \varphi (z_#1)}} 
\newcommand\vtmea[1] {\partial_{z_#1} e^{i(\sqrt{m}-\frac 1 {\sqrt m}) \varphi_1 (z)}}
\newcommand{\vmeo} {e^{i\sqrt{m} \varphi_1 (z)}} 
\newcommand\vme {e^{i\sqrt{m} \varphi (z)}} 
\newcommand\vtme {\partial e^{i(\sqrt{m}-\frac 1 {\sqrt m}) \varphi (z)}}
\newcommand\vm[1] {V(z_{#1}) }
\newcommand\vmn[1] {V(z_{#1}) }
\newcommand\bvm[1] {V_m(\bar z_{#1}) }
\newcommand\vtm[1] { \tilde V_{m}(z_{#1}) }
\newcommand\bhvtm[1] { \hat P_m(\bar z_{#1}) }
\newcommand\hvtm[1] { \hat P_m(z_{#1}) }
\newcommand\bvtm[1] { P_m(\bar z_{#1}) }
\newcommand\holee[1]{e^{ \frac i {\sqrt m} \varphi (\eta_{#1} ) } }
\newcommand\hole[1] {H_m(\eta_#1)}

\newcommand\dcon[2]{\int\hspace{-9pt}\int \put(-8,2){\circle{8}} { \put(-8,2){\circle{5}} }d#1 d#2\, }
\newcommand\dconz[3]{\int\hspace{-9pt}\int_{#1} \put(-8,2){\circle{8}} { \put(-8,2){\circle{5}} }d{#2} d{#3}\, }
\newcommand\dconw[3]{\int\hspace{-9pt}\int_{#1} \put(-10,2){\circle{8}} { \put(-10,2){\circle{5}} }d{#2} d{#3}\, }
\newcommand\dconwn[3]{\int\hspace{-9pt}\int_{#1} \put(-13.8,2){\circle{8}} { \put(-13.8,2){\circle{5}} }d{#2} d{#3}\, }
\newcommand\pump{{\int\!\!\!\!\!\heartsuit}\!\!\!\!\!\!\int }

\newcommand{\CFT } {conformal field theory } 
\newcommand{\etab}{\bar\eta}
\newcommand{\zbar}{\bar z}
\newcommand{\tvphi}{\tilde\varphi}
\newcommand{\rmd}{\mathrm d}

\section{Introduction}

\noi
Ever since Laughlin presented his famous wave function describing the $\nu = 1/m$ quantum Hall (QH) states \cite{laugh}, the construction of trial many-body wave functions has been an important tool to study the properties of the many exotic, strongly correlated states in the QH system\cite{mr,jain}. In a series of recent papers, it was shown how to use CFT to construct a large class of wave functions describing a subset of  the  hierarchical QH ground states as well as their quasiparticle excitations\cite{hansson07,bergh}. Included in this class are the Laughlin wave functions, the  wave functions in the positive Jain series $\nu = 2n/(2pn +1)$, and candidate wave functions for the observed states at $\nu  = 4/11$ and $\nu = 6/17$\cite{pan}.  

The general idea of the CFT construction of QH wave functions is to represent the particles (electrons and quasiholes) of the theory by holomorphic vertex operators in a rational CFT, $V \sim e^{i \sum_j c_j \varphi_j}$ where the $\varphi_j$'s are compact, massless scalar fields. Candidate many-body wave functions of the corresponding QH state are then expressed as correlators, or conformal blocks, of these operators. Depending on the state in question, the wave functions either take the form of a single correlator, or as an antisymmetrized sum over several conformal blocks. The Laughlin states and Moore-Read Pfaffian\cite{mr} are examples of the former, while general states in the Abelian hierarchy fall into the latter class for the following reason: our construction\cite{bergh} for the 
level $n$ hierarchy  states involves $n$ different representations of the electron, $V_{\alpha}(z), \, \alpha=1, ...n$, involving $n$ compact scalar fields $\varphi_\alpha(z)$, with $z$ denoting the position of the particle (we use complex coordinates $z = x + iy$, {\it etc}.); 
in the simplest case, that  of the Jain series, $V_{n}$   can be thought of as a describing a particle in the $n^{th}$ composite fermion Landau level. The sum over correlators was thus necessary to antisymmetrize the wave functions among the different representations $V_{\alpha}$. Correspondingly, there are $n$ independent quasihole operators $H_\alpha (\eta)$, describing quasiholes at positions $\eta$. Candidate wave functions for the ground states at all levels of the hierarchy were thus expressed as linear combinations of conformal blocks containing electron operators $V_{\alpha}$ only; quasihole excitations localized at some points $\eta_i$ are obtained by inserting $\prod_i H_{\alpha}(\eta_i)$ in the correlators.
In the Laughlin case our expressions coincided with the original proposals by Fubini\cite{Fubini},  Moore and Read\cite{mr}, and Wen\cite{wenorig}. 

Although experimentally, QH quasiholes and quasielectrons play a very similar and equally important role, the CFT description
of the quasi{\it electron} required a rather different approach from that of the holes. The reason for this can be intuitively understood as follows:  a quasihole, which   corresponds to a charge deficit, can be created at an arbitrary position $\eta$ just by pushing the incompressible electron liquid away from this point. Mathematically this is achieved  by inserting the local operator $H(\eta)$. The creation of  a quasielectron amounts to  "pulling in" the electron liquid towards the quasielectron position, $\etab$,\footnote{
We shall denote the quasielectron position with $\etab$ rather than $\eta$ for reasons that will be clear later.}
 but this cannot in general be done, since there is a finite probability that there is already an electron at this position. This conflict with the exclusion principle is mathematically manifested in singularities in the electron wave function. More precisely, inserting the inverse of the hole operator $H^{-1}(\eta)$, which does have the correct charge, results in pole terms $\sim 1/(z_i - \eta)$ where $z_i$ is an electron position. The way out of this quandary, which was described in Ref. \onlinecite{hansson07}, is to create the fractional excess charge corresponding to a quasielectron by shrinking the correlation hole around one of the electrons. This amounts to replacing  one of the elecron operators $V_n$ with a modifed electron operator, $P_n$, which turns out to be very closely related to the new electron operator $V_{n+1}$ which appears at the next level in the hierarchy. In this way we could construct quasielectrons in any angular momentum state, and from these it was possible to, by hand, form coherent superpositions describing localized quasielectrons. 

This asymmetry in the description of localized quasiholes versus quasielectrons is somewhat unsatisfactory; an obvious question to ask is whether there exists an operator ${\cal P}(\bar \eta)$ which, when inserted into a correlator of $V_{\alpha}$'s, directly creates a quasielectron localized at $\bar \eta$. In our previous work, it was not clear whether or not this could be done. Although, for the reasons given above, such an operator could not be strictly local, it could still be "quasi-local", \ie the nonlocality is seen only at the magnetic length scale. To clear this out is important; as explained by Moore and Read in Ref. \onlinecite{mr}, we would, on general grounds, expect quasielectrons to enter the theory in a way very similar to quasiholes. In particular,  we would expect the charge and statistics of the quasielectrons to be coded in the properties of this operator, and this could   allow for a better understanding of their braiding phases, and thus their quantum statistics.

In this paper we shall explicitly construct such a {\em quasilocal} operator, $\mathcal P(\etab)$, and show that it has all the general  properties expected for a quasielectron operator. We demonstrate that it can be used to explicitly construct many-quasielectron wave functions of both Abelian and non-Abelian states. In particular, we present explicit two- and four quasielectron wave functions for the Moore-Read Pfaffian state.

The construction of the operator $\mathcal P$ involves two important technical developments. The first concerns the treatment of the background charge needed to neutralize CFT correlators. Our method allows us to directly extract the non-holomorphic Gaussian factors, thus leaving us with purely holomorphic wave functions. The second is related to the precise mathematical meaning of "shrinking the correlation hole around an electron operator". As described above, this  amounts to forming a new local operator by "fusing" an electron operator with the inverse of a quasihole. This is achieved by a generalized normal ordering $(\dots )_{gn}$, which in the simplest case amounts to the conventional normal ordering\cite{gula}, but in general differs from it.

There is a long-standing controversy about whether the composite fermion (CF) wave functions proposed by Jain\cite{jainbook} fit into the original hierarchy scheme of Halperin\cite{halphi} and Haldane\cite{haldhi}. At a superficial level, the two approaches are quite different. The CF scheme predicts explicit wave functions which compare well with numerical simulations, and the CF phenomenology has been very successful in explaining a wide range of experimental data. The hierarchy wave functions, which are expressed as rather complicated multiple integrals, are much harder to deal with, and the phenomenology is much less developed. On the other hand, the newly discovered states that do not fall into the Jain sequences, fit naturally into the hierarchy schemes but are much less straightforward to understand in terms of composite fermions. Also, in a recent work,  Bergholtz and Karlhede proved that when defined on a thin torus or cylinder, the Jain states are precisely condensates of quasiparticles\cite{bekar}. 
Using the formalism developed in this paper, we can explicitly write the full wave functions in the positive Jain series in hierarchical  form, \ie as integrals over multi-quasielectron states with appropriately chosen pseudo wave functions. In our opinion, this shows that the CF states are hierarchical in nature.

The paper is organized as follows. In order to settle notation, and to record some formulae that will be important in the following, we start the next section by summarizing some results from Refs. \onlinecite{hansson07} and \onlinecite{bergh}, on the CFT approach to the hierarchical wave functions. This is followed, in the next subsection, by an outline of a method for handling a homogeneous background charge that will prove to be very useful; some technical details of this discussion are referred to Appendix A. This sets the stage for introducing our construction of a quasi-local quasielectron operator in section \ref{sec:3}, and in order to minimize technicalities we first do this in the simplest case of a single quasielectron in a Laughlin state.
Section \ref{sec:4} begins with  a general discussion of the status of fractional statistics for quasiparticles in the hierarchical quantum Hall States. Then, specializing again to the Laughlin states, we point out the difficulties encountered when naively extending the methods of the previous section to the case of many quasielectrons, and explain how to proceed. An explicit example, that of the two-quasielectron wave function, is worked out in some detail in Appendix \ref{app:B}.
In section \ref{qehier}, which is of a more technical nature and not necessary for understanding the rest of the paper, we introduce the generalized normal ordering mentioned above, and give a definition of the operator $\mathcal P$ which is  appropriate for the more general class of hierarchy states constructed in Ref. \onlinecite{hansson07}. The explicit hierarchical form of the Jain wave functions is derived in section \ref{sec:HiCF}. Finally, section \ref{secMR} discusses the Moore-Read Pfaffian state; here, it is shown how to write down quasilocal quasielectron operators in direct analogy to the Abelian case, and how this leads to explicit two- and four-quasielectron wave functions. We end the paper with some conclusions and a discussion of future directions. A short summary of some of the results of this paper was recently given in Ref. \onlinecite{hhv08}.

\section{The CFT representation of Quantum Hall wave functions}
\label{sec:2}
In order to make this paper reasonably self-contained, we start by summarizing the formalism used in the
CFT description of QH wave functions, as well as recording some formulas that will be used 
later. In the second subsection, we discuss an important technical issue, namely how to introduce a background charge in the CFT correlators in a consistent manner.

\subsection{Operators, currents and correlators} \label{sec:2a}
\noindent
As mentioned in the introduction, it has been shown that a large class of quantum Hall wave functions may be expressed as correlators, or conformal blocks, in certain CFTs. In order to review the basic formalism, we start  with the simplest case, namely the Laughlin states at  $\nu = 1/m=1/(2p+1)$. In this case there are two normal-ordered vertex operators representing the electron and quasihole, respectively,
\be{vo}
V_1 (z) &=& : e^{i \sqrt m \varphi_1(z)} : \\
H(\eta) &=& :e^{ \frac i {\sqrt m} \varphi_1  (\eta) } : \label{hole} \, ,
\ee
and as we shall see in a moment, the wave functions for the ground state and quasihole states can be written as correlation functions of these.
(The normal ordering symbol $:\ \ :$ will be suppressed in the following.
 For simplicity we will also drop the subscripts on fields and vertex operators for the Laughlin states as long as there are no ambiguities.)
Here, $\varphi$ is (the holomorphic part of) a free massless boson field, compactified on a radius $R = \sqrt m$ and defined by the Euclidean action
\be{action}
  S[\varphi] =  \int \rmd^2 x\, \mathcal L =  \frac{1}{8\pi}\int \rmd^2 x\, \partial_\mu \varphi \partial^\mu \varphi \, \
\ee
with $d^2x=dx \, dy =\frac{1}{2} d^2z$. 
The action is normalized to imply the (holomorphic) two point function
$
\av{ \varphi (z) \varphi (w) } = - \ln (z - w)   \, .
$
The vertex operators 
satisfy the operator product expansion (OPE) 
\be{verrel}
 e^{ia\varphi(z) }  e^{ib\varphi (w)  }
	&=&  (z-w)^{ab}  e^{i(a+b)\varphi(w)}   \\  
         &+&   \frac  a {a+b}  (z-w)^{ab+1}    \partial_w    e^{i(a+b)\varphi(w)}   + \mathcal O (z-w)^{ab+2}  \nonumber \, .
\ee
Here and in the following, $\partial$ denotes a holomorphic derivative, while $\bar\partial$ is used for anti-holomorphic derivatives. 
Using standard methods\cite{gula},
one can explicitly calculate the relevant correlation functions. In particular,  the ground state Laughlin wave functions at $\nu  = 1/m$ are given by 
 \be{La}
 \Psi_{1/m}(\{ z_i\}) = \av { \prod_{i=1}^N V(z_i) {\cal O}_{bg} } = \prod_{i<j} (z_i - z_j)^m e^{-\sum_i\frac{|z_i|^2}{4\ell^2}}\, , 
 \ee
where ${\cal O}_{bg} $ is a neutralizing background charge to be discussed in more detail below. Similarly, quasihole states are constructed by insertions of the hole operator $H(\eta)$.

The (holomorphic) $U(1)$ charge density current, $ J(z) =\frac i {\sqrt m} \partial  \varphi (z) $, and the corresponding charge operator, ${\cal Q} = \frac 1 {\sqrt m} \frac 1 {2\pi } \oint dz\,  \partial \varphi (z) $, are normalized so that the electron operators \pref{vo} have unit charge, as can be seen from the commutator $[{\cal Q}, V(z)] = V(z)$; similarly, the charge of the hole operator is $1/m$. This $U(1)$ charge is related to the actual electric charge of electrons and quasiparticles, as described in detail in \oncitep{hansson07}
Translational invariance of the action \eqref{action} implies that only charge neutral correlators are non-vanishing, which is why the neutralizing background charge mentioned above is needed. However, there are additional contraints on the correlators, in particular, a correlator involving the divergence of the current has to satisfy the Ward identity \cite{gula}:
\be{WI}
\bar \partial_\omega \langle J(\omega) \Phi_1(z_1)...\Phi_n(z_n) \rangle
&=& \sum_i q_i \delta^2(\omega-z_i) \langle \Phi_1(z_1)...\Phi_n(z_n)\rangle    \nonumber \, ,
\ee
where the $\Phi_i$'s are arbitrary fields with charges $q_i$. 
Thus, we find that $\bar\partial_\omega J(\omega)$ has support only at the points where a charged field is inserted. 
We shall make use of this in section \ref{sec:3}, where we use the Ward identity to construct a quasilocal  operator that has support only at the positions of the electrons. 

The vertex operators introduced above are primary fields, and as discussed by Moore and Read\cite{mr}, the conformal block given by primary fields, such as \pref{La}, yields a "representative" wave function for the QH state in question. In the simplest cases, these are the actual ground states for idealized (singular) Hamiltonians, but in general they are not known to correspond to any potential. 
The importance of the representative wave functions lies instead in their topological properties and in the assumption that they are "adiabatically" connected to the correct ground state. This raises the question of how to modify the representative wave functions without destroying their good topological properties. To answer this, Moore and Read first pointed out that to each primary field there is a "conformal tower" of descendant fields, which have  the same charge and the same braiding statistics. They then conjectured that the actual wave function, for a realistic potential admitting a QH liquid ground state, would lie in the space spanned by correlators of fields in such conformal towers, and thus has the same topological properties as the representative wave function obtained from the primary fields only. The descendants of a primary field are obtained from the OPE of this field with the energy-momentum tensor, which for the scalar fields considered so far is given by $T(z)=\frac 1 2 :(i\partial\varphi(z))^2:$.  
In general, descendants can be expressed as derivatives of the primary field $V(z)$. In the present paper, we will only be concerned with the simplest case, namely first descendants; the first descendant of an arbitrary primary field $\Phi$ is simply a holomorphic derivative,
\be{L-1}
(L_{-1} \Phi)(z) = \frac{1}{2\pi i} \oint_z dw\, T(w)\Phi(z)&=&\partial_z \Phi(z) \, .
\ee
One should, however, note that replacing a primary (electron) operator with a descendant  will also change the angular momentum. Thus to construct a set of trial wave functions at a fixed $L$, we must restore the angular momentum \eg by multiplying with $Z_{cm} = \sum_i z_i$ to the appropriate power. Another, more physical, approach is to create local particle-hole pairs using the operators $H(\eta)$ and ${\cal P} (\bar\eta)$. The precise relationship between these approaches remains to be clarified,
and one should also understand how they relate to the methods developed using composite fermions. \footnote{There a systematic way of improving the wave functions involves increasing the Hilbert space by including excitons, i.e. states with one or more composite fermions excited to a higher (unfilled) effective Landau level, while conserving the total angular momentum. }

Descendant fields are not only important for constructing improved wave functions, but they are in general  also needed  to get a representative wave function in the first place.
In order to illustrate this point, and show how to go beyond the Laughlin states, we end this subsection with an example from the next level in the hierarchy. The simplest state at the second level is the $\nu = 2/5$ composite fermion state in the positive Jain series. 
Describing the state requires two independent electron operators\cite{hansson07}, which, in a suitable basis\footnote{One is free to use representations with other linear combinations of the boson fields or additional auxiliary boson fields, as long as one does not change the correlations of the electron and hole operators, namely $\av{V_{\alpha}(z) V_{\beta}(w)} \sim(z-w)^{K_{\alpha\beta}}$ and $\av{V_{\alpha}(z) H_{\beta}(w)} \sim(z-w)^{\delta_{\alpha\beta}}$ where {\bf K} is the $K$-matrix of the state\cite{wen}. }, may be expressed as
\be{25eop}
V_1(z) &= & e^{ i {\sqrt 3} \varphi_1(z)}  \\
V_2 (z)&= & \partial \hat V_2 = \partial e^{ \frac {2i} { \sqrt 3} \varphi_1(z)  + \frac {5i} {\sqrt 15} \varphi_2 (z)  } \, . \nonumber
\ee
Here, $\varphi_2$ is an additional, independent bosonic field, described by the same action \eqref{action} as $\varphi_1$. 
In the following, primary fields are denoted by $\hat V_j$, while the $V_j$'s denote electron operators and are in general descendent fields. 
The $\nu = 2/5$ ground state is given by\cite{hansson07}
 \be{25wf}
\Psi_{2/5} =  {\cal A}\av{  \prod_{i=1}^{N/2}  V_{1}(z_i)  \prod_{i=1}^{N/2}  V_{2}(z_i)  {\cal O}_{bg} } \, ,
 \ee
where $\cal A$ denotes antisymmetrization over the electron coordiates $\{z_i\}$. In other words, the ground state is not just a single conformal block as in the Laughlin case, but rather an antisymmetrized sum over correlators involving two different representations of the electron. 
In analogy to the composite fermion picture, these two representations can be thought of as corresponding to composite fermions in the first and second CF Landau level, respectively. 
Since our formulation is entirely within the lowest Landau level, what 
distinguishes the two operators $V_1$ and $V_2$ is the presence of the 
derivative, which is naturally interpreted as giving the particle an 
extra "orbital" spin. This is in turn reflected in a shift in the 
relation between flux and particle number when transcribing the wave 
function eq. \eqref{25wf} to the sphere.

There are also two independent hole operators, determined for instance by the conditions\cite{froh},  
$\av{V_{\alpha}(z) H_{\beta}(\eta)} \sim (z-\eta)^{\delta_{\alpha\beta}}$,
\be{qholes}
H_1(\eta)  &=& e^{ (i/\sqrt{3}) \varphi_1(\eta)   - (2i/\sqrt{15}) \varphi_2 (\eta) } \\
H_2 (\eta)&=& e^{i(3/\sqrt{15}) \varphi_2(\eta)} \nonumber  \, ,
\ee
and quasihole states are obtained by inserting these in the correlator \pref{25wf}.
Less dense QH states at level 2 are obtained by modifying the coefficient of $\varphi_2$ (in such a way that $V_2$ remains fermionic). For example, choosing $\hat V_2 (z)= e^{ \frac {2i} { \sqrt 3} \varphi_1 (z) + \frac {11i} {\sqrt{33}} \varphi_2 (z)  } $, one obtains a candidate wave function for $\nu = 4/11$\cite{bergh}. In general, at level $n$ of the hierarchy the ground state wave functions are constructed from $n$ different electron operators, containing $n$ bosonic fields, and with one additional derivative per level. 
In this way, one can construct explicit wave functions for all hierarchical states corresponding to condensates of quasielectrons; explicit expressions for these wave functions are given in Ref. \onlinecite{bergh}.

\subsection {The background charge}
\label{subsec:bgc}
We now discuss a somewhat subtle, yet important, technical question, namely that of the background charge ${\cal O}_{bg}$. The need for such a background charge is most easily understood in the Coulomb-gas formalism where the vertex operators are fields that carry charge with respect to (in general several) $U(1)$ groups, and the background is needed for the total system to be neutral, and thus to satisfy Gauss' law. In open systems, such as a QH droplet with an edge, one can place a single large charge at a very large distance, $Z$, from the other particles and then extract the correlation functions by  carefully taking the limit $Z\rightarrow\infty$\cite{altbackg}.\footnote{
It is not obvious that this method will result in wave functions describing $N$ electrons forming a quantum Hall droplet - \ie a state that is homogenous up to distances of the order of the magnetic length away from the edge. That this is in fact 
the case can be understood by mapping the compactified plane to the sphere, where a homogeneous background charge can be introduced without breaking the conformal symmetry, by coupling it to the Gaussian curvature\cite{gula}. One can then show that the resulting correlators, which by construction describe a homogenous system, reproduce the result from the plane with the {\em ad hoc} prescription of taking a single compensating background point charge to infinity.} 
At a technical level, the disadvantage of this method is that it cannot be used in finite geometries, such as the torus, or for open geometries different from the circular disc. 

A physically more appealing scheme, that also generalizes to other geometries, is due to Moore and Read\cite{mr} who proposed to use a continuous background charge distribution. For the simplest case of the Laughlin state, this is obtained by using 
\be{bgins}
{\cal O}_{bg} = e^{-i\sqrt m \rho_m \int d^2x\, \varphi(z) } \, ,
\ee
where  $\rho_m = \rho_0/m$, and $\rho_0 = eB/2\pi = 1/2\pi\ell^2$ is the density of a filled Landau level. The generalization to higher hierarchy levels involves several bosonic fields and is straightforward. This choice has the additional advantage of producing the correct Gaussian factor\cite{mr} (see below), so the correlators in fact give the full lowest Landau level wave functions.

Using the background charge \pref{bgins}, however, involves some technical problems. A naive calculation will result in integrals of the form $\int d^2 z\, \ln (z - z_i)$, which are not properly defined. There are several ways out of this quandary. The simplest is to consider correlators of full, as opposed to holomorphic, vertex operators, and at the same time use the full field $\varphi (z,\zbar)$ in \pref{bgins}. Then the relevant integrals in the correlators are of the form $\int d^2 z\, \ln |z - z_i| $, which are well defined up to boundary terms. This approach has allowed us to construct hierarchical states on a torus, but the procedure for extracting the wave functions is less straightforward\cite{torus}.  

A second, purely holomorphic,  approach was introduced  in Ref. \onlinecite{hansson07}, where  we replaced the continuous background charge  with a regular lattice, $z_{\vec n} = (n_x   + i n_y )a $,  of $K = k N$ charges, each of magnitude $-m/k$. The lattice spacing, $a$, is related to the magnetic length by $a^2/\ell^2 = 2\pi m /k$, so the limit $k\rightarrow\infty$ corresponds to taking the lattice spacing to zero. 
In Appendix \ref{app:A1} we outline an extension of this method that allows for calculation of correlators involving insertions of the $U(1)$ current and the energy momentum tensor.

We shall now introduce yet another way to treat the homogenous background charge which will prove to be very convenient in the following. The basic idea is to notice that the exponent of \pref{bgins} is linear in $\varphi$, and thus can be included  as part of the quadratic action by a field-independent shift. This method  has the advantage of using only continuum fields, and will also provide a rigorous setting for the construction of the quasielectron operator in the next section. The following discussion will be entirely in the context of the $\nu = 1/m$ Laughlin states, but the generalization to the hierarchical states is straightforward. The ideas we present here are presumably well known to experts in CFTc, but have to our knowledge not been used in the context of QH physics.\footnote{We are in fact indebted to A. A. Zamolodchikov, for pointing out to us that the transformations \pref{rewrite2} and \pref{rewrite} could be of use in our formalism.}

Let us introduce the shifted field  $ \tvphi(z,\zbar) \equiv \varphi(z,\zbar)  - i\pi\rho \zbar z$
and rewrite the full bosonic action according to
\be{rewrite2}
-\mathcal{S}[\varphi] \rightarrow \int d^2x \, \left( \frac 1 {8\pi} \varphi\nabla^2\varphi - i\rho \varphi \right)= \int d^2x \, \left( \frac 1 {2\pi} \tvphi  \partial\bar\partial \tvphi +  \half \pi\rho^2 \zbar z  \right)
\equiv -\mathcal{S}[\tilde{\varphi}] +\int d^2x \, \half \pi\rho^2 \zbar z  \, ,
\ee
where we neglected possible boundary terms. The currents and vertex operators are recast accordingly as
\be{rewrite}
V(z,\zbar) &=& e^{i\sqrt m\varphi(z, \zbar)}   =  e^{i\sqrt m\tvphi(z, \zbar)} e^{-\sqrt m \rho\pi |z|^2}  
\equiv  \tilde V(z,\zbar) \,  e^{-\sqrt m\rho\pi |z|^2}       \\
 J(z) &=& \frac i {\sqrt m} \partial  \varphi (z, \zbar)  = \frac i {\sqrt m} \partial  \tvphi (z,\zbar) - \frac {\pi\rho} {\sqrt m} \zbar  
  \equiv  J_{p}(z) + J_{bg} \, . \nonumber
\ee
From the first line in \eqref{rewrite}, noting that $\pi \rho \sqrt m = 1/2\ell^2$, we see that the exponential factors needed for the QH wave functions are manifest. The second line in fact shows how the $U(1)$ neutrality condition is implemented, in that the current is decomposed into a piece $J_p$ due to the point-like particles, and a piece $J_{bg}$ coming from the background charge. The total charge is obtained by integrating along a contour enclosing the QH droplet,
\be{chcons}
{\cal Q} = \frac 1 {\sqrt m} \frac 1 {2\pi } \oint dz\,  \partial \varphi (z)  = \tilde{\cal Q}  - N
\ee
where we used $\rho/{\sqrt m} = \rho_0/m = N/A$
with $N$ the total number of electrons, and $A$ the area of the droplet. Thus the original charge neutrality condition ${\cal Q} = 0$ now becomes
\be{modcond}
\tilde{\cal Q}   = N \, .
 \ee
 This argument might seem too simple minded since we have been very cavalier about boundary terms, and one might also reasonably ask precisely how the charge neutrality is implemented in the new variables. In Appendix \ref{app:A2} we present  a Hamiltonian analysis, which does not make any assumptions about the boundary, and  which explicitly demonstrates how the $U(1)$ symmetry is implemented on the quantum states. There we also show that the shift \pref{rewrite} can be implemented at the level of {\it holomorphic} vertex operators by
\be{holshift} 
V (z)  &=& e^{i \sqrt m \varphi(z)}   =  \tilde V(z) \,  e^{-\frac {|z|^2} {4\ell^2} }   .
\ee
The relation \eqref{holshift} will be used in the next section. This way of incorporating the background charge in the action is completely consistent with the previously introduced method of regularizing the background charge by introducing a lattice of fluxtubes. The latter is elaborated in Appendix \ref{app:A1}, and the reader may convince herself that both approaches yield the same wave functions.

\section{The quasielectron operator $ \mathcal P(\eta)  $      }
\label{sec:3}

\noindent
With the preliminaries in place, we are ready to address the central issue of this paper -- how to construct an operator that directly creates a quasielectron localized at a specific point $\eta$. 
In order to present the basic ideas of our construction in the simplest possible way, we will first discuss  the Laughlin states. The generalization to the hierarchical states involves several technicalities and will be deferred to Section \ref{qehier}. 
In order to motivate our construction, we start by discussing the qualitative physical picture, before putting it into a formal language. First, recall the case of quasiholes.
As mentioned in section \ref{sec:2}, the one-quasihole wave function can be constructed by inserting a local quasihole operator, $H(\eta)=e^{(i/\sqrt m)\varphi(\eta)}$, into the correlator \eqref{La}:
\be{cfqh}
\Psi_{1qh}^{(L)}  (\eta; z_1\dots z_N) &=& 
    \av{ H(\eta) \vm 1  \dots  \dots    \vm N  {\cal O}_{bg}    } \\ 
&=&  \prod_{j=1}^N (\eta - z_j) \Psi_L (z_1\dots z_N) e^{-\frac{1}{4m\ell^2}| \eta |^2} \, .
\ee
A naive guess for constructing trial wave functions for quasi{\it electrons} would be to insert the inverse operator, $H^{-1}(\eta)$, into the correlator. This operator has the correct charge and conformal dimension -- but it does not yield appropriate fermionic wave functions, as the correlator includes singular factors, such as $\prod_{i}(\eta-z_i)^{-1}$. The origin of the singularities is the Pauli principle, which makes it impossible to create the excess charge at an arbitrary point, as there is a finite probability that there is already an electron at this position. 

However, there is a  well-defined procedure to generate a localized excess charge, namely by shrinking the correlation hole around one of the electrons, thus allowing the electron liquid to become denser. Shrinking the correlation hole can be thought of as attaching an antivortex, and it creates the expected excess charge, $\frac{e}{m}$. As discussed in Ref.\onlinecite{hansson07}, this construction naturally leads to wave functions of quasielectrons in good angular momentum states. However, the excess charge can be located at an arbitrary position $\eta$ by building linear combinations of attaching the antivortex at different electrons, weighted with a gaussian weight.
In doing this, we have to be careful not to introduce an unwanted antiholomorphic dependence on the electron coordinates $w$.
To proceed, we write
 $|w-\eta |^2=|w|^2 -2 w \bar \eta +|\eta |^2+(w\bar \eta-\bar w \eta)$, and note that since the term in the parenthesis is purely imaginary,  it exponentiates to a pure phase. 
Removing this phase, we can thus use  the remaining terms to localize the excess charge in a region of the size of a magnetic length $\ell$ around the point $w=\eta$ by introducing a  factor $e^{-\frac{1}{4m\ell^2}(|w|^2+|\eta|^2-2\bar\eta w)}$. The constant $1/4m$ was picked so that  $e^{-\frac{1}{4m\ell^2} |\eta|^2}$ is the exponential factor appropriate to a particle with charge $q=-e/m$ moving in the magnetic field, and the remaining gaussian term  $e^{-\frac{1}{4m\ell^2}|w|^2}$ is precisely what is needed to eventually produce the correct gaussian factor for the electrons. 

In order to find an operator that performs the desired attachment of the inverse hole, it is useful to recall the Ward identity \eqref{WI}. The divergence of the conserved charge current has only support on charged sources, in our case at the electron and quasihole positions, and it can therefore be used to place $H^{-1}$ at the electron positions. However, to obtain well-defined expressions a regularization procedure is needed. In the simplest case, \ie the Laughlin states,  this regularization is nothing but the usual normal ordering. The hierarchical states at level $n>1$ require a generalized notion of normal ordering, which will be discussed in Section  \ref{qehier}. For the Laughlin states the quasielectron operator can be written in the following form
\be{qpspec}
\mathcal P(\etab) &=& \int d^2w \, e^{-\frac{1}{4m\ell^2}(|w|^2+|\eta|^2-2\bar\eta w)} \left( H^{-1} \, \bar \partial J_p \right)_n (w) 
\ee
where $ J_p(w)=\frac{i}{\sqrt m} \partial \tilde \varphi(w)$  is nonzero only at the electron positions, and does not see the background charge. We will use the notation   $\mathcal P(\etab) $ rather than $\mathcal P(\eta,\etab) $  since  the $\eta$ dependence in \pref{qpspec}  amounts to a trivial $z$-independent normalization. (The reason for choosing this convention will be clear in a later section.) 
We  can now use the Ward identity \pref{WI} to evaluate the  correlators  $\av{J_p(w)V(z_i)}$ to get   delta functions, $\delta^2(w-z_i)$, 
at the electron positions; the $w$-integral in turn gives terms $\sim \left( H ^{-1}V \right)_n(z_i)$,  with normal ordering symbol, $\left(\dots\right)_n$, defined by
\be{mod}
\left( H ^{-1}V \right)_n(z_i) \equiv    \oint_{z_i} \frac {dz} {z - z_i} \, H^{-1}(z)V(z_i) \, ,
\ee
where the contour is close enough to $z_i$ not to enclose any other singularities\cite{gula}. 
For free fields, the normal ordering \pref{mod} coincides (up to an overall constant) with the normal ordering $:\dots :$ used to define the vertex operators in \pref{vo}.

That $\mathcal P(\etab)$ as defined above is an eigenstate of the total charge follows directly from the commutator, $ [{\cal Q}, H^{-1}(\eta) ] = -\frac 1 m H^{-1}(\eta)$, while the quasilocal nature of the charge is revealed by studying the commutator with the localized charge operator  
\be{co}
{\cal Q}(\eta; \epsilon) = \frac 1 {\sqrt m} \frac 1 {2\pi } \oint_\eta dz\,  \partial \varphi   (z) ,
\ee
 where the contour is a circle of radius $\epsilon$ around $\eta$. We get $ [{\cal Q}(\eta; \epsilon), \mathcal P(\etab)  ] \approx - \frac 1 m \mathcal P(\etab) $ up to an exponentially small correction for $\epsilon \gg \ell$.

Since $J_p$ is expressed in the shifted field $\tilde\varphi$, it is convenient to use the relations \pref{rewrite} to express the original vertex operators, $V$, in terms of the $\tilde V$'s, with the understanding that the expectation values are now taken using the action 
 ${\cal S}[\tilde \varphi]$ \pref{rewrite2}, and  with the modified charge neutrality condition \pref{modcond}. In this way we directly calculate the polynomial part, $\Psi_{1qp}^{(hol)}  (\etab; z_1\dots z_N) $,  of the full wave function, 

\be{holwf}
\Psi_{1qe}^{(hol)}  (\etab; z_1\dots z_N) =      \av{   {\mathcal P}(\etab) {\vtm 1 }  \dots  \dots   \vtm N    } 
\ee
with
\be{holqpop}
\mathcal P(\etab)    =  \int d^2w \,  e^{-\frac{1 }{4m\ell^2} (|\eta|^2-2\bar\eta w)  } \left( \tilde H^{-1} \, \bar \partial J_p \right)_n(w) \, .
\ee
The formula \pref{holqpop} and its generalization \pref{fidef} is one of the main results of this paper.

It was shown in \oncitec{hansson07}  that the CF wave function for a single quasielectron  localized at the origin, in the $\nu = 1/m$ Laughlin state is given by
\be{cfqp}
\Psi_{1qe}^{(CF)}  (\eta=0; \vec r_1\dots \vec r_N) &=& 
    \mathcal{A}\{ e^{- |z_1|^2 /4m\ell^2}  \av{ P(z_1)  \vm 2  \dots  \dots    \vm N  {\cal O}_{bg}    } \}\\
&=&  e^{-\sum_i  |z_i|^2 /4\ell^2} \sum_i (-1)^i  \pr j k {i} (z_j-z_k)^m 
\partial_i  \prod_{l \neq i}(z_l-z_i)^{m-1} \nonumber\\
&=&e^{-\sum_i  |z_i|^2 /4\ell^2}  \av{   {\mathcal P}(\etab=0) {\vtm 1 }  \dots  \dots   \vtm N    }  \nonumber
\ee
where  the operator $ P(z) $ is defined as
\be{qpo}
 P (z) = \left( H^{-1} V\right)_n(z)=          \vtme .
\ee
Thus, inserting the nonlocal operator $\mathcal{P}(\bar\eta)$ can be written as a sum over all electron positions $z_i$ of the local operator $P(z_i)$. Multiple insertions will naturally lead to multiple sums over (different) electron positions. A detailed calculation of the two-quasielectron wave function can be found in App. \ref{app:B}. 
In the earlier construction in \oncitec{hansson07} we were to neglect the non-holomorphic terms $\sim \zbar_i$ coming from the derivative acting on the exponential part of the correlation function. This prescription was introduced {\em ad hoc} in order to get wave functions in the lowest Landau level. Using instead the quasielectron operator \eqref{holqpop} located at the origin to evaluate \eqref{holwf}, we reproduce the holomorphic part of \pref{cfqp} exactly.
In  particular, the derivatives will act only on the polynomial part, so we automatically get holomorphic wave functions without invoking any {\em ad hoc} rule.
In the following we shall, unless stated otherwise, use the shifted fields \eqref{rewrite}, and for simplicity of notation we shall suppress the tildes and write $V$ instead of $\tilde V$.

At this point we should stress a few important  points. 
\begin{enumerate}
\item The derivative occurring in \pref{cfqp} is a direct consequence of "fusing" the operators $H^{-1}$ and $V$ using the standard normal ordering prescription defined by \pref{mod}, which amounts to extracting the leading non-singular term in the OPE.  There is however no a priori reason why we should use this prescription. The only formal requirement on the fusion is that the resulting local operator has the same $U(1)$ charge and the same conformal dimension as the original composite operator. The first condition follows from charge conservation, 
and the second from the composition properties of the (orbital) spin. The latter has a geometric meaning in that it gives rise to Berry phases on curved manifolds, and in particular determines the shift on the sphere which is a topological invariant\cite{QHspin}.
In the hierarchical states where there are several $U(1)$ symmetries, conservation of the charges is needed to get correct charge and (mutual) statistics of the quasiparticles. 
\item From \pref{cfqp} we see that  the final expression for the  electronic  wave function is in terms of correlators involving  primary fields $\hat V$ of the rational CFT and descendants of the form $L^n_{-1} \hat V$ ; as we shall see below, the same will be true for the multi-quasielectron case. This structure is important for two reasons. First, it guarantees that the wave functions are analytic as they should be. The second point relates to the discussion following Eq.\pref{L-1} -- correlators containing descendants have the same braiding properties as those with the corresponding primary fields.
\item We do not expect that inserting $\mathcal P(\etab_1) \mathcal P(\etab_2) $ will  describe a state of two localized quasielectrons, since the operator $P(z)$ in  \pref{qpo} is anyonic, as can be readily seen from the OPE $P(z_i)P(z_j)\sim (z_i-z_j)^{(m-1)^2/m}$. 
Below we shall explain how to supply the phase factors  that are needed to obtain well defined electronic wave functions. 
\end{enumerate}
It turns out that these points are connected, and they are all relevant for finding a $\mathcal P$ operator that can be used at any level in the hierarchy. We will return to this, essentially technical, point later in the paper. 
Finally, we comment that the scheme presented here is manifestly in 
the disk geometry. It would obviously be of interest to carry out this 
construction, including the background charge, in finite geometries. We 
have so far not done this for the sphere; for some recent progress on 
the torus, see Ref.\onlinecite{torus}.
In the following section we discuss a question of principal importance: How do we construct localized states of several quasielectrons, and how do we determine their statistics? 

\section{Several quasiparticles and anyon statistics }
\label{sec:4}

\noindent
One of the most interesting aspects of quantum Hall quasiparticles is their fractional statistics. While the anyonic statistics of Laughlin quasiholes is very well understood, the situation is less clear for quasielectrons in the Laughlin state as well as for any quasiparticle in a general hierarchy state. This section is devoted to a discussion of these issues,  and in particular the statistics of quasielectrons.
We begin with a brief overview of what is known, and what is believed, about the statistics of the quasiparticles in the Abelian hierarchy states. We then turn to a more detailed discussion of fractional statistics within the CFT approach, and
the specific problems we encounter when seeking to generalize the formalism of the previous section to  many-quasielectron excitations, both of the Laughlin states and the more general hierarchical states. 

\subsection{Abelian fractional statistics of Quantum Hall hierarchy quasiparticles} 
We shall here briefly review what is theoretically known about the statistics of the quasiparticles in the Abelian hierarchical states, and discuss in turn holes in the Laughlin states, holes in hierarchical states, and finally quasielectron states. 
\subsubsection{Laughlin quasiholes}
\label{sssec:1}
The anyonic nature of  the Laughlin quasiholes is very well understood. The original calculation of the pertinent Berry phase by Arovas \etal \cite{arovas} confirmed the assertions made earlier by Halperin\cite{halperin84} based on a specific ansatz for the hierarchical wave functions, and later it was realized that the fractional statistics could also be encoded in an effective topological low energy theory of the Chern Simons type\cite{cstheory}. 
The calculation of the Berry  phase assumed that the QH liquid was incompressible, which in turn was proven by Laughlin using the plasma analogy\cite{lauiprange}. Alternatively, one can directly use the plasma analogy to compute the normalization factor $N(\eta_1, \dots ,\eta_n)$ of a state with $n$ widely separated quasiparticles, from which the exchange Berry phase can be extracted. In both cases, the arguments crucially depend on the  plasma analogy, which is applicable only when the holomorphic part of the wave function is of Jastrow form.  The fractional statistics of the Laughlin holes has also been verified by numerical simulations on small systems\cite{nsim1,nsim2}.

Here we should make a technical comment that will be important in the subsequent discussion of the CFT approach: For the Berry phase to be uniquely defined, the wave function must be a single valued function of the adiabatic parameters. In the present context, this translates into the condition that the electronic wave functions must be single valued functions of the holomorphic hole coordinates. If the normalization constant of the electronic wave function for example contains a "monodromy" factor $(\eta_1 - \eta_2)^{\vartheta/\pi}$, the statistical phase pertinent to the exchange of the two quasiholes, will be the sum of  $\vartheta$ coming from  the monodromy, and the Berry phase. 

In addition to the explicit calculations, there is also a very general gauge argument due to Kivelson and Rocek, that ties the fractional statistics of a Laughlin hole to its fractional charge\cite{kivroc}. It goes as follows: At the center of a Laughlin hole the electron density is zero, and thus we can imagine that it supports a thin tube of unit flux -- in fact this was how Laughlin originally argued for the existence of a fractionally charged quasihole. Now imagine transporting another quasihole around the one with a unit flux at the center. Since the quasihole has charge $q = e/m$, it will pick up an Aharonov-Bohm phase equal $2\pi/m$. But  the system is made up only  of charge $-e$ electrons  all moving  entirely in the flux free region, so we can appeal to  the Byers-Yang theorem \cite{BY}   to conclude that the wave function has to be single valued if the flux equals an integer number of elementary flux quanta.   Consequently there must be another phase to cancel the fractional part of the Aharonov-Bohm pase, and this is precisely the fractional statistics phase $\theta$ which is thus equal to $\pi/m$ (the factor of 2 is because a path corresponding to one particle encircling the other is equivalent to two exchanges).  This means that the essentially non-local phenomenon of fractional statistics is closely related to the local phenomenon of fractional charge. The latter is both easier to understand theoretically and simpler to measure in the laboratory\cite{fcexp}.

\subsubsection{Quasiholes in hierarchical states}
Unfortunately the Laughlin states are among the very few where the Berry phases related to quasiparticle exchange can be calculated analytically without additional assumptions. The other important class are the Halperin $(n_1, n_2,m)$ states\cite{halphi}  that describe multilayer systems. In all cases the wave function is a single product of Jastrow factors, which means that the powerful plasma analogy may be applied. 
	For states that are not of this type, the arguments for fractional statistics are much weaker. In the original hierarchy proposal by Halperin, the claim was based on the analysis of the hierarchical wave functions,  which however are too complicted to allow for an explicit calculation of Berry phases. In a later paper Read analyzed the statistics of the hierarchical states in a more precise way, but had to make a "orthogonality postulate", that was made plausible but was not proven\cite{read90}.
Another line of arguments is based on effective Chern-Simons theories, but unlike the Laughlin case, these cannot be derived from the microscopic theory. The third line of argument, which we shall scrutinize more closely below, is based on CFT.

What, if any, are the implications of the Kivelson-Rocek (KR) argument in this case? First we should note that the argument is certainly not always applicable. Take for instance the charge $q=e/4$ quasiholes in the Pfaffian state, which do not obey the expected $\pi/4$ Abelian statistics, but are  known to have non-Abelian statistics. 
The difficulty lies in the fact that the electron density does not vanish at the position of the hole, as can be seen from the corresponding OPE. 
The same holds for the $q=e/5$ quasiholes in the $2/5$ state, so we cannot directly use the KR argument. There is, however, another line of argumentation that strongly suggests that, in the absence of degeneracies, the elementary quasiholes and quasielectrons do obey $p\pi/5$ Abelian fractional statistics for some integer $p$. First we note that according to Laughlin's original gauge argument one can always form a hole with charge $q=\nu e$ by inserting a thin flux line. For such holes the KR argument holds and we can conclude that the statistics is Abelian with a phase $\nu\pi$. Alternatively we note that the Berry phase calculation for the Laughlin quasihole of Arovas \etal is valid for any incompressible state\cite{arovas}.
Next assume that the Laughlin hole can split into two or more fundamental holes, in this case two charge $q=e/5$ elementary holes. Then we can use  a general argument, due to Thouless and Wu, which on the basis of locality suggests that the statistical phase accumulated when a particle encircles a cluster of particles is just the sum of the phases corresponding to encircling the constituents\cite{thwu}.  In CFT language this is reflected in the Abelian fusion rules of the underlying $U(1)$ current algebras. The clustering conditions do not uniquely define the statistical angles, but do give the constraints $\theta_{11}=\theta_{22}$ and 
$\theta_{12} + \theta_{21} = \pi / 5  $. Using the hole operators in Eq. \eqref{qholes}, and noting that the combination $H_1H_2$ corresponds to a Laughlin hole, we find $\theta_{11}=\theta_{22}=3\pi/5 $ and $\theta_{12}=\theta_{21}-2\pi/5 $, which is consistent with the above conditions.
\footnote{But, and this is very important, this conclusion is only valid provided there are no degeneracies in the multi-hole states, \ie the state is completely determined by the position of the quasiparticles. This is not true for the Pfaffian state, and this is precisely what opens the possibility for non-Abelian statistics.  In CFT language, the fusion rules are now more complicated since there is an underlying su(2) curent algebra, which allows for two different splitting channels of the Laughlin hole. } 
If we would further demand that also $H_1H_1$ and $H_2H_2$ should have the same statistics as a Laughlin hole, the solution follows uniquely from the clustering conditions. In an early paper, Su used this stronger condition to get $\theta = 3\pi/5$, but without considering the possibility of two non-equivalent holes\cite{su}.
Although we used the $2/5$ state as an example, we believe that the above types of arguments would also constrain the possible statistical angles of the fundamental quasiparticles in a  general hierarchical state. 

\subsubsection{Quasielectron states}

The status of the quasielectrons is less clear, even for the Laughlin states. First there are two rather different proposals for the Laughlin quasiparticle. 
The first is Laughlins original proposal,
\be{laqe}
\Psi_{1qe}^{(L)}  (\etab; \vec r_1\dots \vec r_N) =  \prod_{i =1}^N (\partial_i - \etab) \Psi_L (z_i\dots z_N)
\ee
which is closely related to the corresponding Laughlin quasihole\cite{lauiprange}. The other is the composite fermion wave function \pref{cfqp}. In neither case is there any simple argument even for the fractional charge. (The gauge argument which works beautifully for the quasihole is not applicable since it yields a wave function that has components in the second Landau level, and has to be projected.) There is however convincing numerical evidence for the charge to equal $-e/m$. The situation concerning the fractional statistics is less clear. Extensive calculations by Kj{\o}nsberg and Leinaas\cite{nsim2}, and by Jeon, Graham, and Jain\cite{nsim3}, have shown that the two-quasielectron wave functions of the Laughlin and Jain type differ substantially in that the latter exhibits a clearly defined statistical angle for well separated quasielectrons, while the former does not. It can of course still be true that the Laughlin quasielectrons have the expected statistics for much larger separation, but that remains to be shown. 

Turning to the hierarchical states, the general arguments based on Chern-Simons theories or the explicit form of the hierarchical wave functions, can be made also for quasielectrons, and are subject to the same type of criticism. It should be clear that the Kivelson-Rocek argument is not applicable in this case, but by using the  Thouless-Wu argument and comparing a quasihole taken around a quasielectron with that of a quasihole taken around another  quasihole, we can conclude that the {\em mutual} fractional statistics between quasielectrons and quasiholes only differs from that of quasiholes by a sign. Then by noting that having a quasielectron go around a quasielectron - quasihole pair is a trivial operation, we conclude that the statistics of a quasielectron equals that of a quasihole with the same charge. So although one should keep in mind that there is no plasma analogy, and thus no conclusive analytical calculation,  the arguments for the fractional statistics of the quasielectrons are nevertheless quite convincing,

\subsection{Quasiparticle statistics in the CFT approach}
\label{sec:4A}

\noi
We now turn to the CFT approach to QH wave functions, and again start, in the first two paragraphs, by discussing the quasihole states, before addressing quasielectrons in the remainder of this section. 

\subsubsection{Laughlin  quasiholes}

\noi
Before addressing the question of two Laughlin quasielectrons, we remind ourselves of the corresponding case of two quasiholes. The pertinent correlators are of the form
\be{2qh}
\av { H(\eta_1)H(\eta_2) V(z_1)\dots V(z_N) } \sim (\eta_1 - \eta_2)^{\frac 1 m} F_{2h}(\eta_1,\eta_2; z_1\dots z_N) \, ,
\ee
 where $V(z)$ is given by Eq.\pref{vo}, $F_{2h}$ is an analytic function of all its arguments, and $\theta = 1/m$ is the fractional statistics parameter of the quasiholes created by $H$. Referring back to section \ref{sssec:1}, we note that this wave function is not single-valued in the quasihole coordinates, so the statistical phase is no longer given by the Berry phase only. In this case  {\em the full statistical phase is given by the monodromy } of the correlation function, or the conformal block. This in turn implies that the Berry phase has to be zero, which can easily be explicitly verified using the plasma analogy. 

\subsubsection{Quasiholes in hierarchical states}
As a concrete example, let us again look at the $\nu = 2/5$ state which according to \pref{25wf}  can be expressed as a correlator of two types of electron operators. Quasihole excitations of this state are obtained by inserting a number of the charge $q = e/5$ hole operators $H_1$ and $H_2$ given by \pref{qholes}. In parallell with the Laughlin case, there are monodromies $(\eta_1 - \eta_2)^{\theta_{\alpha\beta}}$, with $\theta_{11} = \theta_{22} = 3/5$ and mutual statistics $\theta_{12} = \theta_{21} = -2/5$, which is consistent with the expected fractional statistics of the quasiholes\cite{wen}. Here we should comment on that our trial wave function eq.8 is not a single correlator of the form eq.26, but an antisymmetrized sum of such correlators. 
Since, however, the holes have trivial braidings with the electrons, the monodromies will be the same for all terms in the sum, 
and are thus well defined for the full wave function eq. \eqref{25wf}. 
The same will be true in a general hierarchical state.
This, however, does not constitute a proof of the fractional statistics, since we cannot compute the Berry phase, but simply have to assume that it is zero, just as in the Laughlin case. This assumption of having zero Berry phases related to quasiparticle exchange when the wavefunctions are written as CFT correlators, was first tacitly made in the seminal paper by Moore and Read\cite{mr}. A heuristic argument for why this should be true was given in \oncitec{naywil} and an attempt to construct a generalized plasma analogy using the Coulumb gas formulation of CFT was discussed in Ref. \onlinecite{gunay}. Unfortunately, none of these arguments can, in our view, be considered as proofs of a zero  exchange Berry phase. 
 Recently Read has given an analytic argument for the vanishing of the Berry phases for a class of non-Abelian wave functions that can be expressed as a single  conformal block\cite{read08}. Although quite  general in nature, it is not clear to us whether this kind of argument can be applied also  to the states at higher levels in the Abelian hierarchy, where the  wave functions are sums of conformal blocks.

It is certainly not true that the Berry phases corresponding to {\it any} choice of the quasihole operators vanish. To understand this, recall that
within  the CFT framework the fractional statistics is reflected in the OPE of the hole operators, which is directly related to  the monodromies of their conformal blocks.
For example, for the Laughlin states we have $H(\eta_1)H(\eta_2)\sim (\eta_1 - \eta_2)^{1/m}$, and precisely this factor also enters the full correlation functions involving these operators, giving the monodromy and thus the statistics. 
There is however a large freedom in choosing the quasihole operators without changing their long-distance physical properties such as charge and statistics. Taking the $\nu = 1/3$ Laughlin state as an example we can, instead of taking $H(\eta) = e^{i \varphi(\eta)/ {\sqrt 3} }$, use one of the following operators,
\be{altop}
 H_b(\eta)      &=& e^{i \varphi(\eta)/ {\sqrt 3} -i  \sqrt{5/3 }\chi(\eta)   }  \\ 
 H_f(\eta)     &=&  e^{i \varphi(\eta)/ {\sqrt 3} -i  \sqrt{2/3 }\chi(\eta)   }  \, , \nonumber
 \ee
where $\chi$ is a new free scalar field with the same normalization as $\varphi$. In the two-quasiparticle wave function, this will correspond to the replacements $(\eta_1 - \eta_2)^{1/3} \rightarrow (\eta_1 - \eta_2)^2$ and $(\eta_1 - \eta_2)^{1/3} \rightarrow (\eta_1 - \eta_2)$ respectively, which shows that $H_b$ is a bosonic operator while $H_f$ is fermionic. Since this only corresponds to an overall re-phasing of the electron wave functions, the physics of the quasiholes is clearly unchanged and we have just shuffled the statistical phase from the monodromy to the Berry phase. This freedom in choosing the monodromies of the quasiholes is well known, and was, in a different language, pointed out in an early paper by Halperin\cite{halphi}.
Also,  everything said so far directly generalizes to holes in arbitrary hierarchical states. 

We can now make the conjecture about vanishing Berry phases more precise: Quasihole wave functions that are written in terms of conformal blocks of a collection of hole operators $H$ and  $N$ electron operators $V$, have vanishing Berry phases corresponding to the braiding of the quasiholes for large $N$, as long as all fields in the quasihole vertex operators are also present in the electron operator. In the hierarchy case where there are several electron operators $V_n$ and different hole operators $H_a$, all fields in the $H_a$'s must occur in at least one of the $V_n$'s, and the number of electrons, $N_n$ in all the groups must be large. 

At this point many readers might think that we are stressing some rather obvious points, but we shall see that in the case of quasielectrons things are more complicated than for the quasiholes. Within our CFT scheme, the choice of hole operators will no longer be arbitrary, and we can in general not expect to read the statistics from the monodromy. This will be especially important in the subsequent discussion of the non-Abelian Moore-Read state.

\subsubsection{Two quasielectrons and the quasielectron -  quasihole pair}
\label{sssB3}

All the basic problems related to generalizing the approach in section \ref{sec:3} to many-quasielectron states occur already for the simplest case of two quasielectrons or one quasielectron - quasihole pair in the Laughlin state.  The general underlying problem  was  pointed out already at the end of section  \ref{sec:3} --- when we use \pref{holqpop} to fuse an anyonic inverse hole operator with an electron, the resulting operator $P(z)$ is also anyonic and will introduce an unacceptable non-analytic  behavior in the electron coordinates. To take a specific example, the  quasielectron counterpart to \pref{2qh}, using the operator $\mathcal P$ \pref{holqpop}, becomes
\be{2qpna}
\av { \mathcal P (\etab_1) \mathcal P (\etab_2) V(z_1)\dots V(z_N) }    \sim   \sum_{i<j}(-1)^{i+j}\partial_i\partial_j (z_i - z_j)^{\frac 1 m } F_{2p} (\eta_1,\eta_2; z_1\dots z_N) \, .
\ee
As expected, this wave function is not analytic in the {\em electron} coordinates $z_i$ and does thus not describe a state in the lowest Landau level. 

At a qualitative level this problem is quite easy to deal with. Because of the localization implied by Eq.\pref{qpspec}, the sum is dominated by terms where  $z_i$ and $z_j$ are within a magnetic length away from $\eta_1$ and $\eta_2$ respectively. For widely separated quasiparticles, \ie $|\eta_1 - \eta_2|\gg \ell $, we can thus make the replacement $(z_i - z_j) \rightarrow (\eta_1 - \eta_2) $, to get a function which is analytic in the $z_i$'s. Note that the analytic function $F_{2p} $ contains factors $\prod_{k\neq i,j} (z_i - z_k)$ which will not be modified, and it is these factors that determine the fractional charge. 
Although most likely correct for widely separated quasielectrons, this treatment is unsatisfactory for two reasons. First it is hard to estimate the errors involved when $\eta_1$ and $\eta_2$ approach each other. In particular this means that such wave functions will not easily accommodate the dense quasiparticle condensates that are the building blocks of the hierarchy, as explained in the next section. Secondly, since the space of states spanned by the correlators of the operators $V$, $H$, and $\mathcal P$ is not holomorphic, it means that the operator $\mathcal P$ cannot easily be thought of as  a quasilocal counterpart to the local holomorphic vertex operators $V$ and $H$, and this removes much of the formal appeal of our approach. 

The difficulties related to the factor $(z_i - z_j)^{\frac 1 m }$ on the right hand side of \pref{2qpna} were disucssed already in Ref. \onlinecite{hansson07}, and two alternative approaches were proposed. The first was to introduce by hand a suitable factor $(z_i - z_j)^{n - \frac \theta \pi}$ in the wave function. This is not appropriate here since it would amount to replacing the product $\mathcal P(\etab_1) P(\etab_2) $ by a new, essentially non-local,  operator that directly creates two quasielectrons. The other approach was  to change the inverse hole operator $H^{-1}$ to become bosonic, or alternatively fermionic, so that the resulting $P$ operator becomes fermionic or bosonic, respectively. 

It should now be clear how to proceed; the latter approach simply amounts  to using the operators $H_b$ and $H_f$ in \pref{altop} respectively, in the expression for $\mathcal P$.  One should however note that,  as opposed to the case of quasiholes,    
 this modification of the operator does not result in a  mere re-phasing of the electronic wave functions. For instance, using $H_b$, the factor $(z_i - z_j)^{\frac 1 3 }  $ in \pref{2qpna} is replaced by the holomorphic factor $(z_i - z_j)^2$ which amounts to changing both the phase and modulus of the wave function.  In Appendix \ref{app:B} we present an explicit calculation of the two-quasielectron wave function in this basis, along with further technical details.
 
Using a bosonic or fermionic form of the inverse hole operator in $\mathcal P$, the resulting wave functions will be holomorphic in the quasielectron coordinates, so  the fractional statistics will reside entirely in the Berry phase. 
Remember that in the case of quasiholes,  the fractional statistics could be read off directly from the monodromies of the anyonic operators, while introducing the auxiliary fields yielded wave functions that were holomorphic in the quasihole coordinates with the statistics "hidden" in the Berry phase. We now propose that the same is true for the quasielectrons, \ie that their fractional statistics can be obtained directly from the monodromies of the anyonic operators, $H^{-1}$, which do not contain the extra bosonic field $\chi$. 
This amounts to a rather natural extension of the corresponding assumption about vanishing Berry phases in the case of quasiholes. A note of caution is needed here. Since changing the monodromies of the quasielectrons is not merely a rephasing of the wave function, but changes the correlations between the electrons, using different representations leads to slightly different charge profiles.  This is only a short distance effect; large distance properties, such as charge and statistics, should be insensitive to this. 
We should also mention that in \oncite{hansson07} we presented a calculation of the Berry phase of two Laughlin quasielectrons using a random phase approximation, but it is not clear how to extend this to higher levels in the hierarchy. 
 
Note that changing the statistics of the inverse quasihole is done with help of an uncharged (in terms of electric charge) bosonic field, and in particular there is no homogeneous background charge that has a screening effect on this additional field. 
In order to properly define the correlators, we must introduce a compensating charge also for the field $\chi$.
Formally we can do this by again putting a continuous background charge, but this is quite unnatural from the physics point of view, and it is also formally unsatisfactory in that it gives the wrong exponential factors in the wave functions.  
Alternatively, we can neutralize the correlator by putting a compensating charge at a large distance and take the appropriate limit as explained for instance in Ref. \onlinecite{altbackg}. Neither approach is fully satisfactory, although the latter one complies more with the physical picture of creating a quasielectron together with a compensating quasihole or a charged edge excitation. In particular, if we want to view the background charge to be fixed by the ground state, then it should be kept unchanged for the few-quasiparticle case.\footnote{
 We want to stress that this technical difficulty does not affect the construction of the hierarchy ground states where there is a finite {\em density} of quasielectrons, and where the charge is naturally neutralized by a new constant $U(1)$ charge distribution that contributes to the physical  background electric charge. }  \\

The most natural way to achieve charge neutrality is to directly consider quasielectron - quasihole pairs. Using the original, anyonic, hole operators we have, 
\be{qpqh}
\av { \mathcal P (\etab_1) H (\eta_2) V(z_1)\dots V(z_N) }    \sim   \sum_{i}(-1)^i\partial_i  (z_i - \eta_2)^{-\frac {\theta} \pi}  F_{ph}(\etab_1,\eta_2; z_1\dots z_N) \, .
\ee
which is again not analytic in the electron coordinates, due to the prefactor ($F_{ph}$ is analytic in $\{ z_i \}$). This problem can be resolved using the same ideas as in the above case of two quasielectrons. Either we can by hand make the approximation $(z_i - \eta_2)^{-\frac {\theta} \pi} \approx (\eta_1 - \eta_2)^{-\frac {\theta} \pi} $, or we modify the hole operator according to
\be{modhole}
H = e^{i\frac 1 {\sqrt 3} \varphi_1 }   \rightarrow H_{f}   =  e^{i\frac 1 {\sqrt 3} \varphi_1 -i \frac 2 {\sqrt 6} \chi } \, ,
\ee 
and at the same time use the fermionic inverse hole in the quasielectron operator. In this case we create a quasihole-quasielectron pair which does not violate the charge neutrality condition of the correlator. In order to neutralize the correlator, we could equally well have used a bosonic form of the quasihole. This would however not give  an acceptable wave function, since the OPE $H_b(\eta) V_2(z)\sim 1/(\eta -z)$ has a singular term. In the case of quasielectrons only, or quasiholes only, there is however no problem with using the bosonic representation.

In summary, we believe that the CFT approach to QH wave functions has given strong support to the assignment of fractional statistics that was made earlier on the basis of effective CS theories, reasoning based on hierarchical ans\"atze, and on general topological arguments. We also suggest that our CFT approach to quasielectrons, based on  quasi-local operators, provide arguments  for the fractional statistics of the quasielectrons which are at the same level of rigor as those for the elementary quasiholes in the hierarchy states.

\section{The quasielectron operator $\mathcal P$ and the hierarchy }
\label{qehier}

\noindent
As explained in section \ref{sec:2}, states at level $n$ in the hierarchy are constructed using $n$ electron operators and involve correlators of $n$ scalar fields. Since  there are also $n$ inequivalent hole operators $H_i$, there are $n^2$ different candidates for the quasielectron operator \pref{qpspec}, but, as we shall see, only $n$ of these are non-vanishing. Furthermore, in the important case of  the Jain states, only one of these $n$ operators gives a non-zero result when inserted into antisymmetrized correlators of electron fields. A heuristic argument for this can be given in the language of composite fermions\cite{hansson07}: in hierarchical states that are formed from a number of filled "effective Landau levels" of composite fermions, the only way to form a quasielectron excitation is to put a composite fermion in an unfilled level. In the language of CFT, this amounts to using $H_n$, which creates a hole in the highest effective Landau level (defined by the electron 
operator with the largest number of derivatives, or highest orbital spin), in the construction of $\mathcal P$. 

Even after having specifed which $H_i$ to use in the construction, we cannot directly use the formula \pref{qpspec} to obtain $\mathcal P$. As we shall see, the difficulty has to do with the use of the normal ordered product defined in \pref{mod}, and the resolution lies in substituting this with a "generalized normal ordering procedure". This section is technical in nature and can be omitted by any reader who is willing to accept that one can consistently define a product with the property \pref{genmod}. We now proceed to prove this, starting from the simplest case of the $\nu = 2/5$ Jain state.

\subsection{ The $\nu = 2/5$ state}
The $\nu = 2/5$ Jain state is at the second level in the hierarchy, and we refer back to section \ref{sec:2} for the definition of the operators $V_1 $ and $V_2 = \partial \hat V_2$  (where $V_1$ and $\hat V_2$ are primary  fields), and the hole  operators,  $H_1 $  and  $H_2 $. 

Following the logic of the previous section, we now form a bosonic version of the operator $H_2$ by $H_{2b} = H_2 e^{-(7i/\sqrt{35}) \varphi_3 }$, and  then  insert one or more of  the corresponding  $\mathcal P (\etab) $ operators in a correlator containing an equal number of $V_1$ and $V_2$ operators. Before carrying out the $w$-integrals from the $\mathcal P$ operators, we
 notice that the derivatives $\partial_i$ in the $V_2(z_i)$ operators can all be taken out front, to leave us with a  correlator involving the primary fields  $H_{2b}$, $V_1$ and $\hat V_2$. The terms surviving the normal ordering  \pref{mod} will involve the following structures
\be{trouble}
\left(  H^{-1}_{2b}\hat V_{2}  \right) (z) &= & (\partial e^{-\frac{3i}{\sqrt 15}\varphi_2(z)+\frac{7i}{\sqrt{35}}\varphi_3(z)})   e^{  \frac {2i} { \sqrt 3} \varphi_1(z)  + \frac {5i} {\sqrt{15}} \varphi_2(z)   } \\
\left(  H^{-1}_{2b}\hat V_{1}  \right) (z) &= &  e^{  \frac {3i} { \sqrt 3} \varphi_1(z)  - \frac {3i} {\sqrt{15}} \varphi_2(z)     +   \frac {7i} {\sqrt{35}} \varphi_3(z) }    \label{tr2} \, ,
\ee
as is shown by  identifying terms $\sim (z-w)^0$ in the OPE \pref{verrel}.
Note that the first line cannot be written as a total derivative of a vertex operator, and is thus not the first descendant  of any primary field in the theory. This is a technical observation, but it has consequences for the resulting wave functions. Recalling the discussion following Eq.\pref{L-1}, we know that there are general arguments why it is desirable to construct wave functions from primary fields and their descendants only. Moreover, recall that by introducing the bosonic representation for the quasielectrons, all operators in our theory have conformal dimensions such as to ensure that the correlation functions of {\em primary} fields are analytic. This in turn implies that correlators of descendants of the form $L_{-1}^n \Phi = \partial^n \Phi$ where $\Phi$ is a primary field, are also analytic. The way this works is that correlators corresponding to the individual bosonic fields are {\it not} analytic but give cut singularities of the form $(z_i 
 - z_j)^a$, with $a$ non-integer; these however combine to integer powers in the final product, {\it e.g.} $9/3 + 9/15 + 49/35 = 5$ in \pref{tr2}. Thus terms of the type \pref{trouble}, with the derivatives acting on one of the non-analytic factors only, will potentially generate pole singularities in the wave function.

A second difficulty is that we seemingly have two different types of quasielectrons, corresponding to the two operators \pref{trouble} and \pref{tr2}. This might not seem like a problem, but it is in conflict with well-established results on the $\nu=2/5$ state. In particular, this state is known to be very well described by the composite fermion approach, which predicts only one type of quasielectron; thus one would have hoped to reproduce those results. As shown in \oncite{hansson07}, this is achieved if only the product $\left(  H_{2b}^{-1}\hat V_{2}  \right) $ survives.

It turns out that both these problems can be resolved by a redefinition of the normal ordered product \pref{mod}. This will at the same time put $\left(  H_{2b}^{-1}\hat V_{1}  \right) $ equal to zero and convert the first line in \pref{trouble} into a total derivative of a primary field. We can then show that the same will happen at all levels in the hierarchy: only one normal ordered product survives, and the resulting operator is the first descendant (or holomorphic derivative) of a primary field. 

\subsection{The generalized normal ordering }

\noindent
Given the above qualitative arguments against using the naive normal ordering procedure, let us go on to present a generalized normal ordering which does not suffer from these problems. To this end, consider again the OPE,
\be{opex}
H^{-1}_{2b}(w) \hat V_{2} (z) &\sim& \frac 1 {(w -z)} e^{  \frac {2i} { \sqrt 3} \varphi_1 +\frac {2i} {\sqrt {15}} \varphi_2     +   \frac {7i} {\sqrt {35}} \varphi_3 }  \nonumber\\
&+& \left( -\frac 3 {\sqrt{15}}i\partial\varphi_2(w)+\frac 7 {\sqrt{35}}i \partial \varphi_3(w) \right) e^{  \frac {2i} { \sqrt 3} \varphi_1  + \frac {2i} {\sqrt {15}} \varphi_2     +   \frac {7i} {\sqrt {35}} \varphi_3 } 
+ \mathcal O (w - z) \, .
\ee
As noted in the previous subsection, the standard normal ordering amounts to extracting the finite second term by the contour integral \pref{mod}. Alternatively, we could choose to extract the first term, which is a primary field. This has the same $U(1)$ charge as the composite operator on the left, but not the same conformal dimension. 
If we demand that the surviving term of $H^{-1}_{2b} \hat V_{2}$ should have the same conformal dimension as the composite operator, we must supplement it with an operator of dimension 1. The only dimensionful parameters at hand are the quasielectron position $\etab$ and the magnetic length $\ell$. Thus the only two possibilities are $\bar\eta/\ell^2$ and $\partial_i$, since $1/\ell$ would amount to a non-analytic dependence on the external magnetic field $B\sim \ell^{-2}$. The term proportional to $\bar\eta$ is allowed, but vanishes for a quasielectron at the origin, and can in general be shown to correspond to an edge term as discussed in Ref. \onlinecite{hansson07}. We are thus left with the derivative operator. We  cannot, however, just replace  $H^{-1}(z)$ by $\partial_zH^{-1}(z)$. Instead we recall that the operator $L_{-1}$ \pref{L-1} has conformal dimension one and amounts to a derivative when acting on a primary field $\Phi (z)$. These considerations thus lead us to define a generalized normal ordering by $\left(\dots \right)_{gn}$  by
\be{deffus}
\left(  H^{-1} V  \right)_{gn}  (z) \equiv   \oint_z dy\, T(y)\oint_z dw \,H^{-1}(w) V(z) \equiv  \dconz z {y} w \,T(y) H^{-1}(w) V(z) \, ,
\ee
where the double contour integral is defined by the $y$-contour enclosing the $w$-contour. Note that in the Laughlin case, this is equivalent to the conventional normal ordering used in the previous section.  

\noi
With the prescription \pref{deffus} the quasielectron operator takes the form,
\be{fidef}
\mathcal P(\etab)    &=&  \int d^2w \,  e^{-\frac{1 }{4m\ell^2} (|\eta|^2-2\bar\eta w)  } \left( H^{-1} \, \bar \partial J_p \right)_{gn}(w) \\
&\equiv&\int d^2w \, e^{-\frac{1 }{4m\ell^2} (|\eta|^2-2\bar\eta w)  }   \dconw w {y'} y \,T(y') H^{-1}(y) \, \bar \partial_w J(w) \nonumber \, ,
\ee
which should be compared to equation \pref{holqpop}.  It is now easy to verify that
\be{notrouble}
\left(  H^{-1}_{2b}\hat V_{2}  \right)_{gn} (z) &=&  \partial   e^{  \frac {2i} { \sqrt 3} \varphi_1  + \frac {2i} {\sqrt{15}} \varphi_2     +   \frac {7i} {\sqrt{35}} \varphi_3 } \\
\left(  H^{-1}_{2b}\hat V_{1}  \right)_{gn} (z) &=& 0    \nonumber \, .
\ee
Eq. \pref{notrouble} is exactly the operator which in Ref. \onlinecite{hansson07} was found to reproduce the composite fermion wave functions. Thus, inserting one or more of the operators \pref{fidef} into the the correlator \pref{25wf} yields well behaved quasielectron wave functions of the composite fermion type.

\subsection{General hierarchical states}

\noindent
Let us end this section by outlining how our construction generalizes to an arbitrary level $n$ of the hierarchy. At the $n^{th}$  level one has $n$ electron operators $V_{\alpha}$, $\alpha=1,...,n$, and corresponding hole operators $H_{\alpha}$. The latter can always be chosen so that the only singular term in the OPE is given by  
\be{genrel}
H^{-1}_{\alpha}(z)\hat V_{\beta}(w) \sim  \frac 1{(z-w)^{\delta_{\alpha \beta}} } \,    \hat V_{\alpha + 1}(w)     \, ,
\ee
where $\hat V_{\alpha +1}$ is essentially (\ie up to the additional derivative coming from the normal ordering) the new electron operator that will occur at the next level in the hierarchy.\footnote{
As explained in \oncite{bergh}, the operator $V_{\alpha +1}$ is not unique. Technically speaking, there is a freedom in picking the coefficient of $\varphi_{\alpha+1}$ when bosonizing the hole operator $H_{\alpha}$, since any even-integer braiding statistics of $H_b$ is permitted, cf. Eq.\pref{altop} -- this leads to different $V_{\alpha+1}\sim \partial H_{b\alpha}^{-1}V_{\alpha}$. Physically, these different choices correspond to how closely one can "pack" the quasiparticles in the condensate. For instance, starting from $1/3$, the 'minimal' choice $\sqrt{5/3}$ in Eq.\pref{altop} takes us to 2/5. Choosing the coefficient $\sqrt{5/3 + 2} = \sqrt{11/3}$ instead, gives the less dense state 4/11 etc. The relation \pref{genrel} is independent of this choice.}
When computing correlators containing $\mathcal P(\etab)$ and $V_{\alpha}$'s, a pole from \pref{genrel} is necessary for the $y$-integral in \pref{fidef} to be non-zero; thus one gets a contribution only for $\alpha=\beta$, 
\be{genmod}
\left( H^{-1}_{\alpha}\hat V_{\beta} \right)_{gn}  (z)  = \partial \hat V_{\alpha +1} (z)\, \delta_{\alpha \beta} 
\ee
\ie the antivortices described by $H_{\alpha}^{-1}$, can only be attached to the electrons represented by the operator $V_{\alpha}$.  The hierarchical states constructed in \oncite{bergh} are obtained if at every level $n$ one forms the quasielectrons using the operator $H^{-1}_n$. As discussed in some detail in that paper, there is no a priori reason not to consider states formed by condensates involving the other operators $H_{\alpha}^{-1}$ with $\alpha < n$. For a class of states, including the positive Jain series, inserting such operators results in wave functions which vanish under the anti-symmetrization, but we have no reason to believe that this should be true in general.

\section{Explicit hierarchical form of composite fermion wave functions} 
\label{sec:HiCF}

\noindent
Throughout this paper, we have alternatingly been referring to the hierarchy picture and Jain's composite fermion construction when discussing general filling fractions beyond the Laughlin states. There has been a long-term controversy\cite{read90,jainbook} as to whether these two approaches are equivalent. On the surface, they look rather different: In the Haldane-Halperin hierarchy picture\cite{haldhi,halphi}, any quantum Hall state can give rise to a sequence of daughter states as successive condensates of quasielectrons or quasiholes. The Jain model, on the other hand, describes many of the observed states in terms of non-interacting composite fermions moving in a reduced 'effective' magnetic field and filling an integer number of the corresponding effective Landau levels. The latter has led to the construction of explicit, highly successful trial wave functions, while in the former the wave functions are usually given in an implicit form involving multiple integrals over quasiparticle coordinates.
In this section, we present strong evidence that the two approaches are, indeed, equivalent. More precisely, we show that, starting from the $\nu = 1/m$ Laughlin state, the techniques developed in this paper allow for an explicit calculation of the electronic wave functions describing various quasielectron condensates.  In the simplest cases, this  precisely results in Jain's composite fermion wave functions at the second level. Higher levels are not treated explicitly here, but the procedure naturally generalizes to the entire positive Jain series.

According to the  original proposals\cite{halphi,haldhi}, the hierarchical wave functions at level $n+1$ are coherent superpositions of quasiparticle excitations in the level $n$  parent state. For a condensate of quasielectrons this takes the generic form
\be{hiwf}
\Psi_{n+1} ( \vec r_1\dots \vec r_{N}) =   \int  d^2\vec R_1 \dots   \int  d^2\vec R_M    \, \Phi^\star(\vec R_1 \dots \vec R_M ) \Psi_n (\vec R_1 \dots \vec R_M ; \vec r_1 \dots \vec r_{N}) 
\ee
where $\Phi(\vec R_1 \dots \vec R_M ) $ is a suitably chosen  "pseudo wave function" for the strongly correlated state formed by the quasielectrons when condensing. As stressed by Halperin\cite{halphi}, there is a freedom in choosing the statistics of the quasiparticles, and this will be reflected in the analytic properties of the peudo wave function $\Phi$. 

Using our $\mathcal P$ operator, we can now give an explicit expression for the level 1 multi-quasielectron wave function in a Laughlin state,
\be{multqasi}
 \Psi (\vec R_1 \dots \vec R_M ; \vec r_1 \dots \vec r_{N})  = \av {    \mathcal P_b(\etab_1)  \dots   \mathcal P_b(\etab_M)    V(z_1) \dots V(z_{N}) {\cal O}_{bg}     } 
 \ee
where $\eta = X + iY$. The subscript "$b$" on $\mathcal P_b$ indicates that we have used the bosonic form of the quasielectron operator and thus have to take a fully symmetric pseudo wave function $\Phi$. 

The most natural choice for a symmetric pseudo wave function describing $M$ charge $q = -e/m$ bosons in a magnetic field is 
\be{pseubos}
\Phi_k(\vec R_1 \dots \vec R_M ) = \prod_{i<j}^M (\etab_i - \etab_j)^{2k}  e^{-\frac 1 {4m\ell^2} \sum_{i=1}^M |\eta_i|^2}  \, ,
\ee 
where we recall that by the conventions of section \ref{sec:3}, the polynomial part of quasielectron wave functions is holomorphic in $\bar \eta$.
Now, the only $\etab$-dependence in $\mathcal P(\etab_k)$ is in the exponential factor $e^{-\frac{1}{4m\ell^2} (|\eta_k|^2-2\bar\eta_k w) } $, which will turn into $e^{-\frac{1}{4m\ell^2} (|\eta_k|^2-2\bar\eta_k z_i) } $  after the $w$-integration.\footnote{
This was the reason for including the gaussian factor in the definition of $\mathcal P(\etab_k)$.}
    Combined with the gaussian factor from \pref{pseubos}, this yields for each pair ($z_i$, $\bar \eta_k$),
\be{llldel}
\delta (\eta_k - z_i) = e^{-\frac 1 {2m\ell^2}( |\eta_k|^2 -   \bar\eta_k z_i )}
\ee
which is nothing but a lowest Landau level delta function. Since the polynomial part of $\Phi_k$ is holomorphic in $\etab$, the integrand in \pref{hiwf} is holomorphic in $\eta$, and thus the integrals can be done trivially because of the delta functions \pref{llldel}. 

Finally, one has to specify the relation between the number of electrons, $N$, and the number of quasielectrons, $M$. This must be related to the background charge, and follows from demanding that the resulting state be homogeneous; in \oncite{bergh} it was shown how this ratio determines the relative size of the background charges for the two fields $\varphi_1$ and $\varphi_2$. For example, taking $k=0$ corresponds to the densest possible daughter state and implies $N = 2M$; because of the $\delta$-functions,  the integrals in \pref{hiwf} can be computed trivially, giving exactly the $2/5$ Jain wave function. The choice $k=1$ gives the constraint $N=3M$ and a trial wave function for the state at filling fraction $4/11$. 
We should mention that there is a freedom in defining the bosonic operators \pref{altop} - we could have chosen a different coefficient of the $\varphi_2$ field that would have given a stronger repulsion between the bosons by providing an extra factor  $(\etab_i - \etab_j)^{2l} $ in the wave function. But this is precisely the factor that is put explicitly in the pseudo wave function $\Phi$, so it only amounts to a reshuffling of Jastrow factors between the pseudo wave function $\Phi$ and the electron wave function $\Psi$ in \pref{hiwf}.
From this is should also be clear that if we choose a fermionic representation $\mathcal P_f$ for the quasielectron operator, we must take a fully anti-symmetric pseudo wave function, which amounts to having odd powers of the Jastrow factor in \pref{pseubos}. With this change we of course obtain similar electronic wave functions as in the bosonic representation, the only difference being that the derivatives do not act on the Jastrow factors coming from the pseudo wave function.

To summarize, we have shown that the hierarchy wave functions at level 1, \ie those that are obtained by condensing quasielectrons on top of a Laughlin state, can be explicitly written in hierarchical form \pref{hiwf}. 
That it has taken quite a while to realize this is presumably because  the quasielectron states  $ \Psi_n (\vec R_1 \dots \vec R_M ; \vec r_1 \dots \vec r_{N})$ that give the Jain states, are different from the ones  originally proposed  by Laughlin, while our 
choice of pseudo wave functions in \pref{hiwf} is precisely the same as suggested by Halperin. The new result, that allows us to explicitly carry out the integrals over the quasielectron positions, is the explicit form of the quasielectron states expressed in terms of correlators involving the pseudo-local operator $\mathcal P$. It is somewhat surprising that although the quasihole wave functions are much simpler, and have been known for a long time, the  quasihole {\em condensates} are much harder to handle. The resulting  integrals are highly non-trivial, and presently we do not know how to evaluate them.   

\section{The Moore-Read Pfaffian state}
\label{secMR}

\noindent
In the previous sections we have constructed a quasilocal quasielectron operator for the Abelian hierarchical states. The concepts and ideas behind the construction are however very general, and can, in particular, be generalized to the proposed non-Abelian states. Of particular interest among these is the Moore-Read (MR) Pfaffian state which is believed to describe the incompressible state  observed at $\nu=5/2$. This claim is based both on extensive numerical calculations\cite{mrnumerics} and the recent observation of charge $e/4$ quasiparticles. 
Theoretically quite a lot is known about this state. In their original paper, Moore and Read proposed to construct the ground state and multi-quasihole states using CFT operators and also derived 
explicit trial wave functions both for the ground state and the 2-quasihole states\cite{mr}. Most importantly, they pointed out that states with  four, or more, quasiholes should exhibit non-Abelian fractional statistics.  
Later Nayak and  Wilczek\cite{naywil} explicitly computed the four-quasihole wave functions, and showed that they had the expected monodromies. They also made some  conjectures about the structure of the general $2n$-quasihole states and gave an argument for the  degeneracy to be $2^{n-1}$ for fixed quasihole positions. Furthermore Greiter, Wen and Wilczek \cite{GWW} showed that the MR wave function is paired, and  the exact ground state of a singular three-body interaction. Later Read and Green developed the analogy between the MR state and a $p$-wave superconductor and showed that the Ising-type topological charge was related to a Majorana fermion localized on the quantized vortices. This analogy was elaborated further in papers by Ivanov\cite{ivanov}, and Stern \etal \cite{stern}. 

Referring back to our previous discussion about fractional statistics, it should be clear that any conclusions about the non-Abelian fractional statistics based on the monodromies of the wave functions depend on an implicit assumption that the Berry phases vanish. Just as in the case of the hierarchical states, this conjecture has not been rigorously proven, but strong arguments in favor have recently been given by Read\cite{read08}. Additional numerical evidence has been given by  Tserkovnyak and Simon\cite{simon} and by Baraban \etal \cite{baraban}. 

In contrast to the quasiholes, the  quasielectron excitations of the MR state have not yet been studied, mainly for the same reasons as for the hierarchical states -- contracting charge in a naive way leads to singularities at the electron coordinates. In the following we will show how to create quasielectron excitations in the MR state using an appropriate version of \pref{fidef}, and we will give explicit expressions for the two simplest examples, the two- and four-quasielectron states.\\

Before we consider the generalization of  \pref{fidef}, we will briefly summarize what is known about the ground state and the quasihole excitations. The  MR state and its quasihole excitations were originally written in terms of correlators of the operators used in the CFT description of the  Ising model. The electron and hole operators are given by
\be{mrop}
V(z) &=& \psi(z) e^{i\sqrt 2 \varphi(z)} \\
H(\eta) &=& \sigma(\eta)e^{\frac{i}{\sqrt 8}\varphi(\eta)} \, , \nonumber
\ee
where $\psi$ is the  Majorana fermion of the Ising model, $\sigma$ the corresponding spin field, and $\varphi$ a scalar field related to the usual $U(1)$  charge.  
The ground state trial wave function is obtained by computing the correlator of  an even number of electron operators:
\begin{align}\label{MRgs}
\Psi_{5/2}=&\langle \prod_{i=1}^N \psi(z_i)\rangle \langle \prod_{i=1}^N e^{i\sqrt{2}\varphi(z_i)}\mathcal{O}_{bg}\rangle
= \Pf \left(\frac{1}{z_i-z_j}\right)\prod_{i<j} (z_i-z_j)^2e^{-\frac{1}{4\ell^2}\sum_{i=1}^N |z_i|^2},
\end{align}
where $\Pf(A)$ denotes the Pfaffian  of an antisymmetric matrix A, defined by the antisymmetrization over the matrix elements,  $\Pf( A) = \mathcal{A} \{ A_{1,2} A_{3,4} \ldots A_{N-1,N} \}$. 

The appearance of the spin operator $\sigma$ in the operator $H(\eta)$  which creates the  charge $e /4 $ quasihole excitations is crucial since the non-trivial fusion of two spin fields, $\sigma\times \sigma= 1+ \psi$,  is the origin of the conjectured non-Abelian statistics of the quasiparticles; 
it is the presence of two different fusion channels that allows for a non trivial degeneracy of the many-quasihole states. 

In the case of only two quasiholes the wave function is unique; only the correlator where the two $\sigma$ fields fuse to the identity is non-vanishing:
\begin{align}\label{2qhmr}
\Psi_{2qh}(\eta_1,\eta_2)&=\langle H(\eta_1)H(\eta_2)\prod_{i=1}^N V(z_i)\rangle\nonumber\\
&=\Pf\left(\frac{(z_i-\eta_1)(z_j-\eta_2)+(i\leftrightarrow j)}{z_i-z_j}\right)\prod_{i<j}(z_i-z_j)^2 e^{-\frac{1}{4\ell^2}\sum_i |z_i|^2}e^{-\frac{1}{16\ell^2} (|\eta_1|^2+|\eta_2|^2)}.
\end{align}
With four quasiholes the situation is  different. There are two possibilities (or "fusion channels") to fuse the four spin operators to an overall identity operator. These give two distinct trial wave functions\cite{naywil}, which are exactly degenerate for the singular three-body interaction referred to above. 

The naive way to generalize the quasielectron operator to the MR state is to take the above defined operators and insert them into Eq. \eqref{fidef} with suitably modified $T(z)$ and $\bar\partial J(w)$. However, this approach has severe problems, basically for the reasons discussed in section \ref{sec:4} -- by attaching a non-Abelian anyon to the electron operators, we  introduce unacceptable branch cuts in the electronic wave function. Analogously to the Abelian case,  there is a way to 'fix' the problem by hand which yields wave functions that look very similar to those for quasiholes (\cf the discussion at the beginning of section \ref{sssB3}). In particular, there is an explicit non-Abelian monodromy when two quasielectrons are braided. We will show later how the quasielectron states can be obtained in a more consistent and appealing way without resorting to any {\em ad hoc } prescriptions. It is however instructive to first discuss the simplified approach since it gives  some additional insight into the problems associated with non-Abelian charges.

We use the same definition of the quasielectron operator that was  introduced for the Abelian hierarchy states, given by
\begin{align}\label{holqpop2}
\mathcal P(\etab)    =  \int d^2w \,  e^{-\frac{1 }{4m\ell^2} (|\eta|^2-2\bar\eta w)  } \left( H^{-1} \, \bar \partial J \right)_{gn}(w) \, 
\end{align}
with the generalized normal ordering
$\left(  H^{-1} V  \right)_{gn}  (z) \equiv   \oint_z dy\, T(y)\oint_z dw \,H^{-1}(w) V(z) $  (for a detailed definition see \eqref{deffus}).
Remember that only the $\varphi$ field is charged, while $\psi$ and $\sigma$  are topological fields. Therefore the charge current can be written as $J(w)=\frac{i}{2}\partial_w \varphi(w)$, and its divergence has support only at the vertex operators $e^{i\sqrt{2}\varphi(z)}$. In analogy to the Abelian case, we want the generalized normal ordering to give the first descendant of the operator $:H^{-1}V:(z)$, therefore we have to choose the energy momentum tensor to be the sum of the ones for each CFT,  $T=T_\psi+T_\varphi$. 
The pertinent inverse hole operator 
\begin{align}\label{Isinginvhole}
H^{-1}(\eta)=\sigma(\eta)e^{-\frac{i}{2\sqrt 2}\varphi(\eta)}\, ,
\end{align}
is manifestly an anyonic operator that encodes the non-Abelian structure. In case of multiple insertions of quasielectron operators we have to specify their fusion channels.  
Nevertheless, in the case of only two quasielectrons the wave function is well defined and unique; inserting \pref{Isinginvhole} into \pref{fidef}  yields,
\begin{align}\label{MR2qe}
\Psi_{2qe}(\eta_1,\eta_2)&=\langle P(\bar\eta_1)P(\bar\eta_2)\prod_{i=1}^N V(z_i)\rangle\nonumber\\
&=\sum_{\alpha<\beta}(-1)^{(\alpha+\beta)} \left[ \partial_\alpha\partial_\beta \Pf\left(\frac{(z_i-z_\alpha)(z_j-z_\beta)+(i \leftrightarrow j)}{z_i-z_j}\right)(z_\alpha-z_\beta)\prod_{\stackrel{ i<j}{i,j\neq \alpha,\beta}} (z_i-z_j)^2 \prod^N_{\stackrel {i=1} {i\neq \alpha,\beta}  }(z_i-z_\alpha)(z_i-z_\beta)\right]\nonumber\\
&\times e^{\bar N Z_{\alpha\beta}/\ell^2}\cosh(\bar \eta(z_\alpha-z_\beta)/16\ell^2)e^{-\frac{1}{4\ell^2}\sum_i |z_i|^2},
\end{align}
where we introduced center of mass and relative coordinates, $Z_{\alpha\beta}=\frac{1}{2}(z_\alpha+z_\beta)$, $z_{\alpha\beta}=(z_\alpha-z_\beta)$, $\bar N=\frac{1}{2}(\bar \eta_1+\bar \eta_2)$ and $\bar \eta=\bar \eta_1-\bar \eta_2$. 
%
\begin{figure}
{\psfig{figure=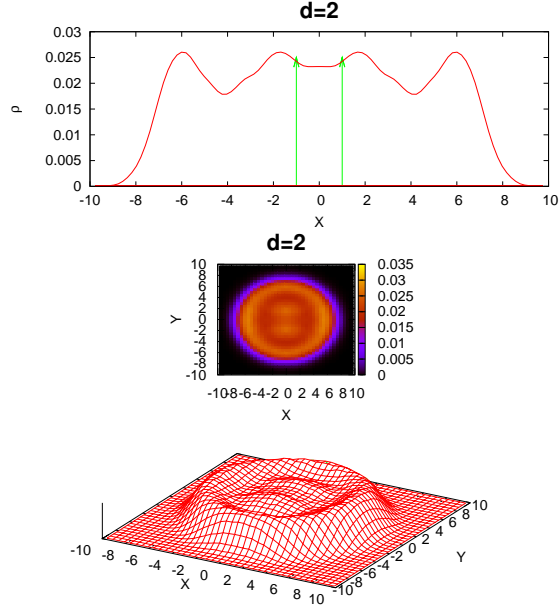, scale=0.6,angle=-0}}
\caption{(Color online) Density profile of the two-quasielectron state \pref{MR2qe} for a droplet of 16 electrons. The quasielectrons are separated by a distance of two magnetic lengths.}
\label{fig1}
\end{figure}
%
%
%
Figure \ref{fig1} illustrates the density profile of this state for two quasielectrons centered around the origin in a circular droplet consisting of 16 electrons. Numerical evidence that the charge of each of the quasielectrons indeed equals $-e/4$, was given in Ref. \onlinecite{hhv08}. 

Note that the computation of the  two-quasielectron wave function for $\nu=5/2$  is completely analogous to the hierarchical states, as two spin fields behave as bosons. This will not be  true any longer in the four-quasielectron case, where the quasielectrons are manifestly non-Abelian. 
Still, we can use the non-Abelian operator to compute the correlator for four quasielectrons and find:
\begin{align}\label{Is4qefix}
\Psi_{4qe}^{(p)}(\eta_1,\ldots, \eta_4) &=\langle \mathcal{P}(\bar\eta_1)   \mathcal{P}(\bar\eta_2)  \mathcal{P}(\bar\eta_3)  \mathcal{P}(\bar\eta_4)   \prod_{i=1}^N V(z_i)\rangle_{(p)}\nonumber\\
&=\sum_{I}(-1)^{\sigma(I)} e^{(\bar\eta_1 z_\alpha+\bar\eta_2z_\beta+\bar\eta_3z_\gamma+\bar\eta_4z_\delta)/8\ell^2} e^{-\frac{1}{4\ell^2}\sum_{i=1}^N |z_i|^2-\frac{1}{4m\ell^2}\sum_{\alpha=1}^4 |\eta_\alpha |^2}\left(\prod_{\alpha\in I}\partial_\alpha \right)\Psi_I^{(p)}(I)\times \Psi_B(I)
\end{align}
where 
$I=\{\alpha, \beta, \gamma,\delta\}$ denotes the set of four (in general not radially ordered) electrons, to which the inverse holes are attached, and $(-1)^{\sigma(I)}$ is the sign obtained by rearranging the electrons from radial order to the one where the electrons of set $I$ are permuted to the left. The fusion channel of the quasielectron pairs at $(\eta_1,\eta_1)$ and $(\eta_3,\eta_4)$ is coded in $p=0,1$, where $p=0$ denotes that both pairs fuse to the identity, while for $p=1$ each pair fuses to a Majorana fermion $\psi$.    
The bosonic contribution is easily computed, 
\begin{align}\label{Psi_B}
\Psi_B(I)=\prod_{\alpha<\beta \in I}(z_\alpha-z_\beta)^{9/8}\prod_{\alpha \in I}\prod_{j\not\in I} (z_\alpha-z_j)^{3/2}\prod_{\stackrel {i<j} {i,j\not\in I }}(z_i-z_j)^2,
\end{align}
and the Ising part can be computed in the same way as in \oncitec{naywil} or using the bosonization techniques detailed in for instance \oncitec{FFN}
\begin{align}\label{Psi_I}
\Psi_I^{(p)}(I)&=(z_\alpha-z_\beta)^{-1/8}(z_\gamma-z_\delta)^{-1/8}\prod_{a\in I}\prod_{j\not\in I}(z_a-z_j)^{-1/2} \left[(1-x)^{1/4}+(-1)^p(1-x)^{-1/4}\right]^{-1/2} \\
&\times \left[(1-x)^{1/4}\psi_{(\alpha, \gamma)(\beta, \delta)}+(-1)^p (1-x)^{-1/4}\psi_{(\alpha, \delta)(\beta, \gamma)}\right] \, ,\nonumber
\end{align}
where
\be{subwf}
\psi_{(\alpha, \beta)(\gamma,\delta)} =\Pf\left(\frac{(z_i-z_\alpha)(z_i-z_\beta)(z_j-z_\gamma)(z_j-z_\delta)+(i\leftrightarrow j)}{z_i-z_j}\right)  ,  \, \,\,i,j\neq \alpha,\beta, \gamma,\delta\nonumber
\ee
and
$x =(z_\alpha-z_\beta)(z_\gamma-z_\delta)/(z_\alpha-z_\delta)(z_\beta-z_\gamma)$ is the anharmonic ratio.

As it stands, \pref{Is4qefix} is clearly not an acceptable wave function because of the branch cuts in the electron coordinates. In analogy with the Abelian hierarchy states, one would think that this problem could be solved by supplementing the hole operator with an auxiliary spin field $\tilde\sigma$ that does not couple to the electrons. As it  turns out, the
overall non-analytic factor can be taken care of by introducing two auxiliary fields,  but there is a relative singular factor between the two independent functions, $\psi_{(\alpha ,\gamma)(\beta, \delta)}$ and $\psi_{(\alpha, \delta)(\beta, \gamma)}$, that we have  not been able to  eliminate within a representation based on the Ising operators. 
Just as discussed in the beginning of section \ref{sssB3}, we  can however simply replace the electron coordinates in the non-Abelian factors by the pertinent quasielectron coordinate.  This yields trial wave functions with a structure that is very similar to the quasihole wave functions derived in \oncitec{naywil}
\begin{align}\label{MR4qehol}
\Psi_{4qe}^{(p)}&=\sum_{I}  e^{(\bar\eta_1 z_\alpha+\bar\eta_2z_\beta+\bar\eta_3z_\gamma+\bar\eta_4z_\delta)/8\ell^2}\left[\sqrt{(\eta_1-\eta_3)(\eta_2-\eta_4)}+(-1)^p \sqrt{(\eta_1-\eta_4)(\eta_2-\eta_3)}\right]^{-1/2}\nonumber\\
&\times \prod_{\alpha\in I} \partial_\alpha\left[\sqrt{(\eta_1-\eta_3)(\eta_2-\eta_4)} \, \psi_{(\alpha, \gamma)(\beta, \delta)}+(-1)^p \sqrt{(\eta_1-\eta_4)(\eta_2-\eta_3)} \, \psi_{(\alpha, \delta)(\beta, \gamma)}\right]\nonumber\\
&\times\prod_{\stackrel {i<j} {i,j\not\in I }}(z_i-z_j)\prod_{i<j} (z_i-z_j) e^{-\frac{1}{4\ell^2}\sum_{i=1}^N |z_i|^2-\frac{1}{16\ell^2}\sum_{\alpha=1}^4 |\eta_\alpha|^2}.
\end{align}
Note that the derivatives act on everything on the right except the exponential factors, and that the last Jastrow factor includes {\it all} electrons in radial ordering.   
The non-Abelian braiding properties are manifest in the wave functions, as \eg  braiding  $\eta_4$ around $\eta_1$ will switch between the two fusion channels.
In analogy with the quasi-hole case discussed above, it is rather natural to conjecture that the non-Abelian part of the fractional statistics is fully captured by the monodromies of the wave functions \pref{MR4qehol}, and that  at most an Abelian phase factor is contributed by the  Berry matrix. 
However, the replacement of the non-Abelian factors is done by hand, and although the final result looks quite reasonable, we are not making a controlled approximation to a well-defined expression.  For this reason we now turn to an alternative description of the MR state that will allow for a consistent description of the quasielectron states in terms of the operator \pref{fidef}.

The basic idea is to use an alternative representation of the MR states which only utilizes scalar fields. This representation is essentially the one used for the 331 Halperin state\cite{halphi} which describes a two layer (or spin 1/2)  state at filling factor $\nu=2$. The crucial difference is that the wave function is antisymmetrized, thus corresponding to only a single layer (or, equivalently, a single spin polarization). The close connection between the MR state and the 331 state was originally pointed out by Ho\cite{ho} and later developed by Cappelli \etal \cite{cappelli}. 

In terms of the two independent, chiral bosonic fields $\varphi(z)$ and $\phi(z)$, the electron and hole operators take the form
\be{qhbos}
V(z) &=& \cos(\phi(z)) e^{i\sqrt 2 \varphi(z)}  \nonumber \\
H_{\pm}(\eta)&=& e^{\pm i \phi(\eta)/2} e^{ \frac{i}{\sqrt{8}} \varphi(z)} \, ,
\ee
where the "charge" field, $\varphi(z)$, is identical to the bosonic field used in \eqref{MRgs}, while the Ising part is replaced by exponentials of the "topological" field, $\phi(z)$\cite{zuber}.

There is an important distinction between the two scalar fields in the way the neutrality condition for the correlators is satisfied. For the charge field $\varphi(z)$ we shall put the usual homogeneous background charge ${\mathcal O}_{bg}$ \pref{bgins}, while there is no such background charge for the topological field. 
Thus, because of the $U(1)$  topological charge neutrality in the $\phi$ field, the quasiholes must be inserted in pairs, $H_+H_-$, and one can show that different orderings span the space of the quasihole wave functions found in Ref. \onlinecite{naywil}. It should be stressed that as long as only quasiholes are considered, the two representations \pref{mrop} and \pref{qhbos} give equivalent electronic wave functions (corresponding to two different basis choices), and it is purely a matter of taste which one to use. 

However, in contrast to the quasihole operators in the Ising representation, we can change the statistics of the operators \eqref{qhbos} at will by introducing two new fields, $\chi_1$ and $\chi_2$, and this will allow for a consistent definition of quasielectrons using \pref{holqpop2}. For instance, by defining 
\begin{align}\label{fermqh}
H_{\pm}&=e^{\frac{i}{\sqrt 8}\varphi}e^{\pm \frac{i}{2}\phi}e^{-i\sqrt{\frac{3}{8}}\chi_1\mp \frac{i}{2}\chi_2},
\end{align}
the holes have fermionic monodromies and their OPE's are given by
\begin{align}
H_\pm(z)H_\pm(w) &\sim (z-w)    \nonumber\\
H_-(z)H_+(w)&\sim (z-w)^0 ,
\end{align}
where only the leading short distance behavior is indicated. We  could equally well have chosen a bosonic representation of the quasiholes, in which case the operators are given by
\begin{align}\label{bosqh}
H_{\pm}&=e^{\frac{i}{\sqrt 8}\varphi}e^{\pm \frac{i}{2}\phi}e^{-i\sqrt{\frac{7}{8}}\chi_1\mp i \frac{\sqrt{3}}{2}\chi_2},
\end{align}
and the OPE's  by
\begin{align}
H_\pm(z)H_\pm(w)&\sim (z-w)^2  \nonumber\\
H_-(z)H_+(w)&\sim (z-w)^0 .
\end{align}
Using these hole operators, the monodromies will be either absent or just a sign, corresponding to bosons and fermions respectively 
(and in the latter case it is also important to remember that the holes are not allowed to be at the same point). 
In the Ising representation, on the other hand, the hole wave functions have non-Abelian monodromies. Thus there must be a corresponding non-Abelian Berry matrix in the 331 representation.

In the following, we will use the fermionic representation of the quasiholes. Using \eqref{fermqh}, we can write two distinct quasielectron operators $\mathcal{P}_{\pm}$:
\begin{align}\label{qebos}
\mathcal{P}_{\pm}(\bar\eta)&= \int d^2 w \, e^{-(|\eta|^2-2 \bar \eta w)/16\ell^2} \left(H_\pm^{-1}\bar \partial J_p\right)_{gn}(w) 
\end{align}
where, as explained in section \ref{sec:3}, the current does not act on the background charge, and the energy momentum tensor in the generalized normal ordering is defined as to give an overall partial derivative. 
\begin{align}
\left(H^{-1}_\pm V \right)_{gn}(z) &\equiv \partial_z \hat V_\pm(z)\hspace{1cm}\mbox{ and}\nonumber\\
\left(H^{-1}_\pm\hat V_+\right)_{gn}(z)&= \left(H^{-1}_\pm\hat V_-\right)_{gn}(z) =0
\end{align}
with
\begin{align}
\hat V_\pm(z)&=e^{i\frac{3}{\sqrt 8}\varphi(z)\pm \frac{i}{2} \phi(z)}e^{i\sqrt{\frac{3}{8}}\chi_1(z)\pm \frac{i}{2}\chi_2(z)}
\end{align}
Note that also the quasielectron operators have to be inserted in pairs, for the neutrality condition to be fullfilled. 
For two quasielectrons at positions $\eta_1$ and $\eta_2$, $\Psi_{2qe}=\langle \mathcal{P}_+(\bar\eta_1)\mathcal{P}_-(\bar\eta_2)\prod_{i=1}^NV(z_j)\rangle$, we exactly reproduce the wave function \eqref{MR2qe} we found using the Ising representation.
Note that choosing the reverse order of the $\mathcal{P}$'s leads to the same wave function because of an overall $\mathbb{Z}_2$ symmetry. Therefore, in the case of four quasielectrons there are only three possibilities of arranging two $H_+$ and two $H_-$, which will be denoted by $\Psi_{4qe}^{(\eta_1,\eta_2)(\eta_3,\eta_4)}$,  $\Psi_{4qe}^{(\eta_1,\eta_3)(\eta_2,\eta_4)}$ and $\Psi_{4qe}^{(\eta_1,\eta_4)(\eta_2,\eta_3)}$. Here, the functions are not labeled by the fusion channel $p=0,1$ \footnote{In fact, they do not correspond to a specific fusion channel but rather a linear combination of the two.}, but by the different orderings of the $\mathcal{P}_+$'s  and $\mathcal{P}_-$'s. The notation is such that the first parenthesis contains all positions of the $\mathcal{P}_+$'s  while the second one those of the $\mathcal{P}_-$'s.  A straightforward calculation yields (the gaussian factors are suppressed)
\begin{align}\label{4qeMR}
\Psi_{4qe}^{(\eta_1,\eta_2)(\eta_3,\eta_4)}&=\sum_{I} e^{(\bar\eta_1 z_\alpha+\bar\eta_2 z_\beta +\bar\eta_3 z_\gamma+\bar\eta_4 z_\delta)/8\ell^2}\partial_\alpha\partial_\beta\partial_\gamma\partial_\delta\nonumber\\
&\times \left[(z_\alpha-z_\beta)(z_\gamma-z_\delta)\psi_{(\alpha, \beta)(\gamma, \delta)}\prod_{\stackrel {i<j} {i,j\not\in I }}(z_i-z_j)\prod_{i<j}(z_i-z_j)\right] \, ,
\end{align}
and similarly for  the other two orderings. As in the quasihole case, these three functions  are not independent, and using the identities given in \oncite{naywil}one can derive the following relation:
\begin{align}
 (z_1-z_2)(z_3-z_4)\psi_{(1,2)(3,4)}&=(z_1-z_3)(z_2-z_4)\psi_{(1,3)(2,4)}-(z_1-z_4)(z_2-z_3)\psi_{(1,4)(2,3)}  \, ,
\end{align}
which shows that  there are two independent four-quasielectron states at fixed positions. Using the 331 representation, computing the correlators for $2n$ quasielectron insertions is trivial. The remaining problem lies in finding the linearly independent functions, and thus the degeneracy.  For $2n$ quasielectrons there are 
$\frac{(2n)!}{2(n!)^2}$ different orderings.  However, using the Pfaffian identities derived in Ref. \onlinecite{naywil} we can argue that only $2^{n-1}$ of these are linearly independent. Thus, we find that the multi-quasielectron states behave in the same way as the multi-quasihole states, namely there are  $2^{n-1}$ independent states for $2n$ quasiparticles at fixed positions.

Using the quasilocal operator \eqref{fidef}, the quasielectrons in the MR state can be described in a way completely analogous to the quasiholes, albeit with the statistics 'hidden' in the Berry matrix. Appropriate linear combinations of the functions \eqref{4qeMR} should have manifest non-Abelian statistics. We want to stress again that  there are no conceptually new problems in the MR case, all difficulties encountered were already present for the hierarchical states. In particular, we believe that our construction can be generalized naturally to other non-Abelian states.

During the completion of the manuscript, we received a preprint of another proposal for quasielectrons in the MR state \cite{domkraft} using Jack polynomials. Comparison of these different methods is work in progress. For Abelian quasiparticles both should give equivalent wave functions. In Ref. \onlinecite{domkraft}, the Abelian quasielectron in  the MR state  is constructed in the same way as the Jain quasielectron\cite{jainbook} in the Laughlin states. On the other hand, using the Abelian quasihole in Eq. \eqref{fidef} yields an Abelian quasielectron that is also completely analogous to the Jain quasielectron in the hierarchical states. Thus, it seems very likely that both methods give the same quasielectron wave function, and preliminary analytical and numerical results confirm this. There is no {\it a priory} reason to believe this to be the case for non-Abelian quasielectrons. However, preliminary results indicate that even in the non-Abelian case the wave functions for a single quasielectron-quasihole pair are identical, though for several such pairs the two methods give different wave functions.

\section{Summary and outlook}

\noindent
The central result of this work is the construction of the quasilocal operator ${\cal P}(\bar \eta)$ that creates localized quasielectron states when inserted into the relevant CFT correlators. Our formalism closely parallels the well-established description of quasiholes in terms of local operators $H(\eta)$ and allows for an explicit construction of many-quasielectron wave functions for all states in the Abelian hierarchy that fall into the class of quasielectron condensates. As a direct consequence of this, we were able to prove that Jain's composite fermion wave functions can be written explicitly in hierarchical form, \ie as  a condensate of quasielectrons of the parent state. Although we have not been able to make any mathematically rigorous statements about the statistics of the quasielectrons, we proposed that the close analogy between correlators involving local hole operators $H(\eta)$ and quasilocal quasielectron operators ${\cal P}(\bar \eta)$, strongly suggests the statistics of the latter, and might eventually allow for a {\em bona fide} analytic calculation. 
Moreover, the form of the quasielectron operator is sufficiently general to be applicable to non-Abelian states. This led us to explicit candidate wave functions for localized multi-quasielectron excitations of the MR Pfaffian state.

The construction of ${\cal P}(\bar \eta)$ also involved a couple of useful technical developments. The first was the introduction of a generalized normal ordering procedure that allowed us to 'fuse' inverse holes with electron operators at a general level of the hierarchy. The other was an alternative way of handling the neutralizing background charge which led to a purely holomorphic CFT description of quantum Hall wave functions.

The results reported here open a number of future directions. The perhaps most exciting is  the possibility of forming condensates of {\it non-Abelian} quasielectrons, which might give rise to a hierarchy of genuinely non-Abelian states not suggested previously. This would involve calculations along the lines of section \ref{sec:HiCF}, but with the MR quasielectrons of section \ref{secMR}. We are presently investigating this problem. Other open questions concern the construction of the operator $\mathcal P(\bar \eta)$ in finite geometries, how to treat quasi{\it hole} condensates along the lines of section \ref{sec:HiCF}, and a better understanding of general quasiparticle - quasihole states, in particular in the MR case.

\bigskip
\noindent
{\bf Acknowledgments} We would like to thank Anders Karlhede for many helpful discussions and in particular for finding an error in an early version of this work.
We also thank A.A. Zamolodchikov for useful discussions, Nicolas Regnault for providing numerical plots, and Eddy Ardonne for useful discussions and a critical reading of the manuscript. Financial support from the Swedish Research Council, the Norwegian Research Council and NordForsk is
gratefully acknowledged.

\begin{appendix}

\section{The background charge}

\subsection{Hamiltonian approach to the background charge}
\label{app:A2}

\noi
This appendix presents the Hamiltonian analysis of the background charge, leading to the fully holomorphic formulation promised in section \ref{subsec:bgc}.
The analysis is most conveniently carried out in real time. We thus start from a canonical two-dimensional massless boson defined on a circle of 
 with circumference $L$, and later translate the results into the Euclidian plane. The field operator is first expanded in normal modes as 
\be{phicyl}
\varphi (x, t) = \sum_n e^{\frac {2\pi i n} L x} \varphi_n (t) \, ,
\ee 
	The action $\tilde S_M = \int dx dt\,   \tilde {\mathcal L}_M$, where $\tilde{\mathcal L}_M$ is the Minkowski form of the shifted Lagrangian given in \pref{rewrite2},  gives rise to  the Hamiltonian
\be{ham}
H = \frac {2\pi} {L} \sum_n \{\pi_n\pi_{-n} + \frac {n^2} 4 \varphi_n\varphi_{-n} + 2\pi \rho e^{\frac {4\pi i} L t} i \varphi_0 (t)    \}
\ee
with the momenta $\pi_n = \frac L {4\pi} \dot\varphi_{-n}$ , and the canonical commutation relations,
\be{comrel}
[\varphi_n , \pi_m ] =  i\delta_{n,m} \, 
\ee
Note that the background charge term translates into a complex term in the Hamiltonian. The resulting theory is not unitary, which in the original complex Euclidian formulation is reflected in factors $e^{-\frac 1 {4\ell^2} |z|^2}$ which amounts to an exponential decay in the radial "time" direction. 

We now concentrate on the zero mode $\{ \varphi_0, \pi_0\}$ which satisfies the Hamiltonian equations,
\be{hameq}
\dot \varphi_0 (t) &=& \frac {4\pi} L \pi_0 \\
\dot \pi_0 (t) &=& - i \frac {4\pi^2\rho} L   e^{\frac {4\pi} L it}   \nonumber \, ,   
\ee
with the solution
\be{hamsol}
\pi_0 (t) &=&  \pi_{0c} -  \pi\rho  e^{\frac {4\pi} L it}  \\
\varphi_0 (t) &=&  \varphi_{0c} +      \frac {4\pi} L t\pi_{0c} + i \pi\rho e^{\frac {4\pi} L it}   \, , \nonumber
\ee
with $\pi_{0c} $ and $ \varphi_{0c}$ being time-independent. 
Note that in the presence of the background charge, \ie $\rho\ne 0$, the momentum $\pi_0$ is no longer a constant of motion. Also, as expected, the other modes are not affected by the background charge. 

After going back to the complex plane using the transformation, 
\be{z}
z = e^{2\pi(\tau - ix)/L} 
\ee
where $\tau = i t = (L/4\pi) \ln \zbar z$, 
the solution \pref{hamsol} reads,
\be{hamsolc}
\pi_0 (\zbar z) &=& \pi_{0c} -  \pi\rho  \zbar z  \\
\varphi_0 (\zbar z) &=& \varphi_{0c}  -i  \pi_{0c}\ln \zbar z+ i \pi\rho \zbar z \, .  \nonumber
\ee

We can now follow standard procedure\footnote{
See \eg chapter  9 in \oncitep{gula}}
to address the question of charge neutrality, which only affects the zero mode. In the absence of the background charge, we define a time independent vacuum as an eigenstate of the operator  $\pi_0$,
\be{vac}
\pi_0 \ket {\beta} = \beta \ket {\beta} \, .
\ee
Since a vertex operator $V_a$ has the zero mode part $ e^{ia\varphi_0} $ which shifts the $\pi_0$ eigenvalue according to
\be{shift}
e^{ia\varphi_0 } \ket\beta = \ket {\beta + a} \, ,
\ee
the insertion of a number of vertex operators with charges $a_i$ evolves an original state $\ket\beta$ to the state $\ket{\beta + \sum_i a_i}$, and finally the orthogonality condition $\bracket \beta {\beta '} = \delta (\beta - \beta ')$ implies the charge conservation condition $\sum_i a_i =0$ for the correlator $\bra\beta \prod_i V_{a_i} \ket \beta$ to be non zero.

When the background charge in included, $\pi_0$ is no longer a constant of motion, but the eigenvalue $\beta$ is evolving in time according to \pref{hamsolc}, 
\be{timeev}
\ket {\beta(\zbar z)} = \pi\rho \zbar z \ket {\beta (0)} \, .
\ee
This shifts the neutrality condition to 
\be{newcond}
\sum_i a_i = \pi\rho \zbar z = \pi \rho R^2 = \rho A
\ee
where $A$ is the area of the circular droplet. This is precisely the neutrality condition given in the main text. 

We can now give the fully holomorphic formulation promised in the text. With a slight misuse of notation, we define the "holomorphic" field as 
\be{phihol}
\varphi (z) &=& \varphi_{0c}  - i \pi_{0c} \ln z + \frac{ i\pi\rho} 2 \zbar z  + i \sum_{n\neq 0} \frac 1 n a_n z^{-n} \nonumber\\
 \tilde \varphi (z) &=& \varphi(z) - \frac{ i\pi\rho} 2 \zbar z\, ,
\ee
where we  included the full zero mode $\varphi_{0c}$, but only half of the non-holomorphic term $\sim \bar z z$. As usual, this means that $\varphi (z,\zbar) = \varphi (z) + \bar\varphi (\zbar) - \varphi_{0c}$ because of the double counting of the zero mode. 
Finally, note that while $\varphi $ is not strictly holomorphic, the shifted field $\tilde \varphi$ is, and this is the field that eventually enters in all correlators, as explained in the main text.

\subsection{Flux tube regularization of the background charge} 
\label{app:A1}
The regularization of the background charge introduced in \oncite{hansson07} amounts to replacing the homogeneous background charge \eqref{bgins} by a regularized version $\mathcal{O}_{rbg}$, where  the total flux is distributed on $K = kN$  fractional flux tubes of strength $\delta \phi=\frac m k \phi_{0}$. We shall only treat the case of a $\nu = 1/m$ Laughlin state, described by the vertex operators $V(z_j)=e^{i\sqrt{m} \varphi(z_j)}$,  but the  generalization to the other states considered in this paper should be straightforward.  For simplicity, we consider the flux tubes to be at positions $z_{\vec{n}}$ on a square lattice, with lattice constant $a$
\be{regins}
{\cal O}_{bg} \rightarrow  {\cal O}_{rbg} = \,    : \prod_{\vec n} V_b( z_{\vec n}  ): \,  =\, : \prod_{\vec n} e^{-i \frac {a^2} {\sqrt m 2\pi \ell^2}  \varphi (z_{\vec n} ) } :\, = \,  :\prod_{\vec n} e^{-i \frac {\sqrt m} k \varphi (z_{\vec n} ) }: \, .
\ee
In this formulation, all operators are primary fields of a CFT, albeit with a different radius than the original CFT, namely $R^2=m\rightarrow R^2=\frac {k^2} m$. This allows for a rigorous definition of the CFT correlators. 

In \oncite{hansson07} it was furthermore shown that in the limit of vanishing lattice constant the correlators involving the $V$'s  are of the form
\be{regcorre}
\av{\prod_i V_i(z_i) {\cal O}_{rbg}  } =   e^{i\sum_i \Phi(z_i, \bar z_i) }  f(\{z_i\})     \prod_{i,\vec n}  |z_i - z_{\vec n} | ^{ -\frac {m}{k} }
\ee
where $ f(z_i)$  is a regular holomorphic function of $\{ z_i\}$, and the singular phase factor is given by
$
e^{i \Phi(z, \bar z) } =  \prod_{\vec n} \left[  { (\bar z - \bar z_{\vec n} )} /{( z - z_{\vec n} )}     \right]^{\frac {m}{2k} } \, .
$
Taking $ a/\ell \rightarrow 0 $, and neglecting boundary terms, we have
$
   \prod_{\vec{n}}  |z - z_{\vec{n}} | ^{ -\frac {m}{k} } \rightarrow e^{- \frac{ |z|^2 }  {4 \ell^2} }
$, 
so in this limit we can recover the QH wave function in the symmetric gauge up to a well defined, albeit singular, gauge transformation. 

Difficulties however occur when we try to calculate correlators involving operators other than vertex operators. In the construction of the quasielectron operator we encounter both the current operator $J(z)=\frac{i}{\sqrt m} \partial \varphi(z)$ and the energy momentum tensor $T(z)$. Considering the OPE of of the background charge with $J(w)$, we find $J(z) e^{-i\frac{\sqrt{m}}{k}\varphi( z_{\vec n } )} \sim \frac{1}{k(z-z_{\vec n})}e^{-i\frac{\sqrt{m}}{k}\varphi(z_{\vec{n}})}$. This is by no means surprising, as the flux tubes carry fractional charge. Although this does not cause a problem for the definition of the quasielectron operator as $(H^{-1}V_b(z_{\vec n}))=0$, it is still preferable to make the background 'invisible' by introducing a covariant derivative: 
\be{covdiv}
iD\varphi (z)  = i\partial \varphi (z) - \frac{\sqrt{m}}{k}\sum_{\vec n} \frac 1 {z - z_{\vec n} },
\ee
and redefining the charge current $ \mathcal{J}(w)=\frac{i}{\sqrt m}\partial \varphi(w)-\frac{1}{k}\sum_{\vec n}\frac{1}{z-z_{\vec n}}$. 
There is however an even more important reason to introduce this derivative, namely to make the action of $T(z)$ well-defined. $T(z)$ is not only used in the definition of $\mathcal{P}$, but is also needed  to construct the descendants of the electron operators that are important for constructing realistic wave functions.  
It is obvious from Eq. \eqref{regcorre} that  introducing first (or higher) descendants of the electron operators will yield spurious pole singularities at the lattice points, which will prevent us from taking the continuum limit. However, this can be cured by using \eqref{covdiv} to redefine the energy momentum tensor
\be{newdef}
T(z) \rightarrow {\cal T} (z) =\frac 1 2  :(i D \varphi (z) )^2: \, .
\ee
The redefinition amounts to replacing 
\be{replace}
\partial_{j} V_j(z_j)\rightarrow \left[ \partial_{j}+\frac{m}{k}\sum_{\vec n} \frac{1}{z_j-z_{\vec n}}\right] V_j(z_j) \, ,
\ee
in the definition of descendant operators such as \pref{qpo}. The additional term precisely cancels the poles we encounter by letting the derivative act on the background charge contribution in Eq. \eqref{regcorre}, and the continuum limit is well defined. Note that in this limit, the derivatives do not act on the gaussian factors and the polynomial part of the wave functions are purely holomorphic. 

In \oncite{hansson07} it was remarked that in the continuum, the rule that the holomorphic derivatives only act on the polynomial part of the wave functions could be implemented by the substitution 
\be{adhoc}
\partial_i \rightarrow \partial_i + \frac 1 {4\ell^2} \zbar \, ,
\ee
and it  is instructive to see how this prescription emerges from \pref{covdiv} by the use of \pref{replace}. For simplicity we consider a correlator with only one insertion of a descendant operator \pref{replace}, and derive the large $k$ limit as,
\be{regcorre2}
&&\av{\prod_{i\ne j} V(z_i) \left[\partial_{j}+\frac{m}{k}\sum_{\vec n} \frac{1}{z_j-z_{\vec n}}\right] V(z_j) {\cal O}_{rbg}  } \\
 &=&   \prod_{i, \vec n} \left( \frac { z_i - z_{\vec n} } { \zbar_i -  \zbar_{\vec n} } \right) ^{-\frac {m}{2k} }  
 \left[\partial_{j}+\frac{m}{2k}\sum_{\vec n} \frac{1}{z_j-z_{\vec n}}\right] f(\{z_i\})     \prod_{i,\vec n}  |z_i - z_{\vec n} | ^{ -\frac {m}{k} } \\
 &\rightarrow&  e^{i\sum_i \Phi(z_i, \bar z_i) } \left[ \partial_j + \frac 1 {4\ell^2} \zbar  \right]  f(\{z_i\})  e^{-\frac 1 {4\ell^2} \sum_i |z_i|^2}
\ee
so we see that the prescription \pref{adhoc} is correctly reproduced in the continuum limit.
In deriving the limit, we used
\be{sumapp}
\lim_{k\rightarrow\infty}  \frac m {2k} \sum_{\vec n} \frac 1 {z - z_{\vec n} } =  \frac m {2ka^2} \int d^2 r'\, \frac 1 {z - z'}\,  \bar\partial'  \zbar'  
= \frac 1 {4\ell^2} \zbar \, .
\ee
where the last equality follows since $ka^2 = 2\pi m \ell^2$ and $\bar\partial \frac 1 z = \pi \delta^2(\vec r)$.

In summary, we have presented a consistent way to calculate correlators involving both vertex operators and insertions of currents and the energy-momentum tensor within the flux tube regularization of the background charge.

\section{The two-quasielectron wave function }
\label{app:B}
Here we give the explicit form of the two-quasielectron wave function in the Laughlin state and also explain some subtleties in the derivation.  In the following, we consider two quasielectrons at position $\bar \eta_{\pm}=\bar N\pm \bar \eta/2$  in the fermionic representation $\mathcal{P}_f$, \ie we use the fermionic hole \eqref{altop} in the operator \eqref{holqpop}:
\be{2qp}
\Psi_{2qp}&=&\langle \mathcal P_f(\bar\eta_+)\mathcal P_f(\bar\eta_-)\prod_{j=1}^N V(z_j)\rangle \,
\ee
Taking proper care to satisfy the charge neutrality condition (see section \ref{sec:4A}) and  also carefully evaluating all possible operator contractions, \pref{2qp} can now be evaluated to become a double sum, where the quasielectron at position $\eta_+$ attaches to the electron at $z_i$ and the other quasielectron attaches to the electron at $z_j$:
\be{}
\Psi_{2qp}&=&\sum_{i\neq j}  e^{-\frac{1}{4m\ell^2}(|\eta_+|^2-2\bar\eta_+ z_i)}e^{-\frac{1}{4m\ell^2}(|\eta_-|^2-2\bar\eta_- z_j)}\\
&&\langle \dconwn {z_i\;} {y_1} {y_1'} T(y_1)H_f^{-1}(y_1')\dconwn {z_j\;} {y_2} {y_2'} T(y_2)H_f^{-1}(y_2')\prod_{k=1}^NV_1(z_k)\rangle\nonumber\\
&=&-\sum_{i<j}(-1)^{i+j}e^{-\frac{1}{4m\ell^2}(|\eta_+|^2-2\bar\eta_+ z_i+|\eta_-|^2-2\bar\eta_- z_j)}\av{P_f(z_i)P_f(z_j)\prod_{k}\!^{(i,j)}V_1(z_k)}\nonumber \\
&&+\sum_{i>j}(-1)^{i+j}e^{-\frac{1}{4m\ell^2}(|\eta_+|^2-2\bar\eta_+ z_j+|\eta_-|^2-2\bar\eta_- z_i)}\av{P_f(z_i)P_f(z_j)\prod_{k}\!^{(i,j)}V_1(z_k)}\nonumber\\
&=&-\sum_{i<j}(-1)^{i+j}e^{-\frac{1}{4m\ell^2}(|\eta_+|^2+|\eta_-|^2)}(e^{\frac{1}{2m\ell^2}(\bar\eta_+ z_i+\bar\eta_- z_j)}-e^{\frac{1}{2m\ell^2}(\bar\eta_+ z_j+\bar\eta_- z_i)})\av{P_f(z_i)P_f(z_j)\prod_{k}\!^{(i,j)}V_1(z_k)} \nonumber
\ee
with $P_f(z)=\partial e^{i\frac{m-1}{\sqrt m}\varphi_1(z)+i\sqrt{\frac{m-1}{ m}}\chi(z)}$. The contribution where both quasielectrons attach to the same electron vanishes. A detailed calculation can be found below. 
The sign in the second equality depends on whether or not $V(z_i)$ must be commuted through $V(z_j)$. After changing to the relative and center of mass  coordinates, $z_{ij}=z_i-z_j$ and $Z_{ij}=\frac{1}{2}(z_i+z_j)$, we find 
\be{2pqres}
\Psi_{2qp}&=&\tilde{ \mathcal{N}}_2\sum_{i<j} (-1)^{i+j} e^{-\frac{1}{2m\ell^2}(|\bar N|^2-2\bar N Z_{ij})}e^{-\frac{1}{8m\ell^2}|\bar \eta|^2}  \sinh(\frac{\bar \eta z_{ij}}{4m\ell^2})  \langle  P_f(z_i)P_f(z_j){\prod_k }^{(i,j)} V(z_k)\rangle
\ee

Up to an $\eta_{\pm}$-dependent normalization constant, this expression is  identical   to the corresponding wave function given in \oncitep{hansson07} Choosing the bosonic instead of the fermionic representation gives a  very similar wave function, where $\sinh(\frac{\bar \eta z_{ij}}{4m\ell^2})  \langle  P_f(z_i)P_f(z_j)\ldots\rangle$ is replaced by $ \cosh(\frac{\bar \eta z_{ij}}{4m\ell^2})  \langle  P_b(z_i)P_b(z_j)\ldots\rangle$, and $P_b(z)=\partial e^{i\frac{2}{\sqrt 3}\varphi_1(z)+i\frac{5}{\sqrt {15}}\chi(z)}$. These two wave functions differ in their behaviour when the two quasielectrons get near to each other, but the symmetry properties and the leading exponential factor for large distances are the same. For further discussion, see Refs. \onlinecite{hansson07} and \onlinecite{qestat}.

We now explain in some more detail how \pref{2pqres} is derived. 
The quasielectron operator acts in principle on all fields with vorticity, in particular also on other quasielectrons as well as on holes. While this is no problem for the single-quasielectron wavefunction, there might be unwanted contributions when several quasielectrons or holes are present. We must thus calculate all possible contributions explicitly and show that only contractions with the electron operators survive. The argument is in fact rather simple. Only (sufficiently) singular OPEs with the inverse hole will have a non-vanishing contour integral. As the singularities in the OPE depend on the charge, we find that the quasielectron operator effectively only acts on charge one operators. 

Consider a general charged operator $\phi(z)$ with electric charge $\alpha$. Because $\bar \partial_w J(w)$ has support at every charged operator, inserting $\mathcal{P}_f$ into a correlator that involves $\phi$ gives a contribution
\be{att}
\langle \mathcal{P}_f(\eta)\phi(z)\ldots\rangle&=&\alpha \langle \int d^2 w \, e^{-\frac{1}{4m\ell^2}(|\eta|^2-2\bar \eta z)} \oint_z dy' \,T(y') \oint_z dy \,  H^{-1}_f(y) \phi(z)\ldots \rangle + \ldots \, ,
\ee
and only the term where $\mathcal{P}_f$ attaches to $\phi$ is written explicitly. The contour integral is only non zero if the OPE of $\phi$ with $H_f^{-1}$ yields a single pole: $H_f^{-1}(y) \phi(z)\sim (y-z)^{-1}:H_f^{-1}\phi:(z)$.  This shows immediately that there is no contribution of a quasielectron attaching to another quasielectron, as $H^{-1}_f(y)H^{-1}_f(z)\sim (y-z)^1 :H^{-2}_f:(z)$ is regular. The same is valid if we try to attach to a modified electron operator, $P_f(z)$. However, it does not rule out the possibility that a quasielectron can attach to a quasihole, as their OPE yields a single pole. In this case it is the second contour integral that vanishes as $H^{-1}_f(y)H_f(z)\sim (y-z)^{-1} {\bf 1}$, \ie they combine to the identity operator. The second contour integral amounts to taking the first descendant of the enclosed operator, and the first descendant of the identity operator vanishes, $\oint dy \, T(y) {\bf 1}=0$. Therefore, the quasielectron operator $\mathcal{P}_f$ indeed only attaches to electron operators, and different quasielectron operators cannot attach to the same electron. 
Note that if we had used an anyonic $\mathcal{P}$ operator, the contour integrals in these expressions would not have been well defined because of the cut in $H(\eta)H(\xi)\sim  (\eta-\xi)^{1/m}$. Note that all arguments in this appendix are equally valid if we had chosen the bosonic representation instead.

\end{appendix}

\end{document}